\documentclass[a4paper, 10pt, twocolumn]{article}
\usepackage{geometry} 
\usepackage{multicol}		
\usepackage{titlesec}
\titleformat{\section}{\normalfont\large\bfseries}{\thesection}{1em}{}
\titleformat{\subsection}{\normalfont\normalsize\bfseries}{\thesubsection}{1em}{}

\titlespacing\section{0pt}{12pt plus 3pt minus 3pt}{3pt plus 1pt minus 1pt}
\titlespacing\subsection{0pt}{10pt plus 3pt minus 3pt}{3pt plus 1pt minus 1pt}
\titlespacing\subsubsection{0pt}{8pt plus 3pt minus 3pt}{3pt plus 1pt minus 1pt}

\def\abstract{\flushleft{\bf Abstract.}  } 

\usepackage[numbers,round,sort&compress]{natbib}
\usepackage{microtype}
\usepackage{mathtools,amsfonts,dsfont,amsmath,amssymb}
\usepackage{tabularx}
\usepackage{xcolor}		
\definecolor{mycolor}{RGB}{74,168,142}
\usepackage[colorlinks=true, linkcolor=mycolor, urlcolor=mycolor, citecolor=mycolor, anchorcolor=mycolor]{hyperref}

\newcommand*{\tens}[1]{\underline{#1}} 

\newcommand*{\vcenteredhbox}[1]{\begin{tabular}{@{}c@{}}#1\end{tabular}} 

\newcommand\blfootnote[1]{
	\begingroup
	\renewcommand\thefootnote{}\footnote{#1}%
	\addtocounter{footnote}{-1}%
	\endgroup
}

\usepackage[]{changes}
\begin{document} 

\title{\Large \bf Modeling the effects of environmental and perceptual uncertainty using deterministic reinforcement learning dynamics with partial observability}
\author{Wolfram Barfuss$^{1,2}$ and Richard P. Mann$^{2}$} \date{}
\maketitle
\thispagestyle{empty}

\begin{abstract}
Assessing the systemic effects of uncertainty that arises from agents' partial observation of the true states of the world is critical for understanding a wide range of scenarios, from navigation and foraging behavior to the provision of renewable resources and public infrastructures. Yet, previous modeling work on agent learning and decision-making either lacks a systematic way to describe this source of uncertainty or puts the focus on obtaining optimal policies using complex models of the world that would impose an unrealistically high cognitive demand on real agents.
In this work we aim to efficiently describe the emergent behavior of biologically plausible and parsimonious learning agents faced with partially observable worlds. Therefore we derive and present deterministic reinforcement learning dynamics where the agents observe the true state of the environment only partially.
We showcase the broad applicability of our dynamics across different classes of partially observable agent-environment systems.
We find that partial observability creates unintuitive benefits in a number of specific contexts, pointing the way to further research on a general understanding of such effects.
For instance, partially observant agents can learn better outcomes faster, in a more stable way and even overcome social dilemmas.
Furthermore, our method allows the application of dynamical systems theory to partially observable multiagent leaning. In this regard we find
the emergence of catastrophic limit cycles, a critical slowing down of the learning processes between reward regimes and the separation of the learning dynamics into fast and slow directions, all caused by partial observability.
Therefore, the presented dynamics have the potential to become a formal, yet practical, lightweight and robust tool for researchers in biology, social science and machine learning to systematically investigate the effects of interacting partially observant agents.
\end{abstract}

\blfootnote{Published in \href{https://doi.org/10.1103/PhysRevE.105.034409}{Phys. Rev. E \textbf{105}, 034409} \newline
	$^1$University of Tübingen, GER \ $^2$University of Leeds, UK \newline
	Contact: wolfram.barfuss@gmail.com $|$ \today}

\section{Introduction}
\label{sec:Intro}

We do not observe the world as it is, but instead as our limited sensory and cognitive apparatus perceives it. There are always elements of the world that are hidden from us, such as the detailed physical state of our environment and the internal states of other agents. As such uncertainty is a fundamental feature of life.
To be more specific, we might not know what will happen (\textit{stochastic uncertainty}), what currently is (\textit{state uncertainty}) and what others are going to do (\textit{strategic uncertainty}), among other forms of uncertainty \citep{kochenderfer2015decision,halpern2017reasoning,marchau2019decision}.
In common with other animals, we must learn and make decisions amid this uncertainty using the limited cognitive resources available to us. So must everybody else.

Given the cognitive demands of fully integrating all sources of uncertainty when learning from experience and making decisions, real agents must employ methods of bounded rationally \cite{simon1997models} that use cognitive resources efficiently to obtain acceptable solutions in a timely manner \cite{griffiths2015rational}. As such, evolutionary game theory \citep{HofbauerSigmund1998} takes into account \textit{strategic uncertainty} by assuming that other agents are not perfectly rational but instead by allowing agents to adapt to each other sequentially, with relatively-successful strategies being reinforced and less successful strategies selected against.
Tools and methods from evolutionary game theory have also been used successfully to formally study the dynamics of multiagent reinforcement learning \citep{BloembergenEtAl2015,barfuss2020towards}.
\citet{BoergersSarin1997} established the formal relationship between the learning behavior of one of the most basic reinforcement learning schemes, Cross learning \citep{Cross1973}, and the replicator dynamics of evolutionary game theory. Since then, this approach of evolutionary reinforcement learning dynamics has been extended to stateless $Q$-learning \citep{TuylsEtAl2003,SatoCrutchfield2003}, regret-minimization \citep{KlosEtAl2010} and temporal-difference learning \citep{BarfussEtAl2019}, as well as discrete-time dynamics \citep{GallaFarmer2013}, continuous strategy spaces \citep{Galstyan2013} and extensive-form games \citep{PanozzoEtAl2014}.
This learning dynamic approach offers a formal, lightweight and deterministically reproducible way to gain improved, descriptive insights into the emerging multiagent learning behavior.

Apart from strategic uncertainty, representing \textit{stochastic uncertainty}, i.e., uncertainty about what will happen in the form of probabilistic events within the environment, requires foremost  the presence of an environment.
Recent years have seen a growing interest to move evolutionary and learning dynamics in stateless games to changing environments. Here, the term environment can mean external fluctuations \citep{assaf2013cooperation,ashcroft2014fixation}, a varying population density \citep{hauert2006evolutionary,gokhale2016eco}, spatial network structure \citep{gracia2013cooperation,szolnoki2018environmental}, or coupled systems out of evolutionary and environmental dynamics.
Coupled systems may further be categorized into those with continuous environmental state spaces \cite{tavoni2012survival,weitz2016oscillating,chen2018punishment,tilman2020evolutionary,WangFu2020eco} or discrete ones \citep{hilbe2018evolution,BarfussEtAl2019,hauert2019asymmetric,su2019evolutionary}.
We will focus on learning dynamics in stochastic games \citep{hilbe2018evolution,BarfussEtAl2019} which encode stochastic uncertainty via action-depended transition probabilities between environmental states.

However, all dynamics discussed so far are either applicable only to stateless environments, assume that agents do not tailor their response to the current environmental state, or if they do, assume that agents observe the true states of the environment perfectly. Yet, often in real-world settings state observations are noisy and incomplete. Thus, they are lacking a systematic way to describe interacting agents under \textit{state uncertainty}.

In this work, we relax the assumption of perfect observations and introduce deterministic reinforcement learning dynamics for partially observable environments.
With the derived dynamics we are able to study the idealized reinforcement learning behavior in a wide range of environmental classes, from partially observable Markov decision processes \citep[POMDPs,][]{Spaan2012}, decentralized \mbox{(Dec-)}POMDPs \citep{OliehoekAmato2016}, and fully general partially observable stochastic games \citep{HansenEtAl2004}.

Note that while a great deal of works on partially observable decision domains is of normative nature, ours is descriptive.
For the normative agenda, agents are often enriched with, e.g., generative models and belief-state representations \citep{Spaan2012,OliehoekAmato2016},  abstractions \citep{SuttonEtAl2006} or predictive state representations \citep{littman2001predictive} in order to learn optimal policies in partially observable decision domains. Also the economic value of signal is often studied by asking how fully rational agents optimally deal with a specific form of state uncertainty \citep{bagh2020economic}.
However, such techniques can become computationally extremely expensive \citep{LochSingh1998}.
It is unlikely that biological agents perform those elaborate calculations \citep{gigerenzer2011heuristic}
and the focus on unboundedly rational game equilibria lacks a dynamic perspective \citep{papadimitriou2019game} making it unable to answer which equilibrium (of the often many) the agents select.

Instead, this work takes a dynamical systems perspective on individual learning agents employing the widely occurring principle of temporal-difference reinforcement learning \citep{sutton1988learning} in which the agents simply treat their observations as if they were the true states of the environment. Temporal-difference learning is not only a computational technique \citep{SuttonBarto2018}, it also occurs in biological agents through the dopamine reward-prediction error signal \citep{schultz1997neural,DayanNiv2008}.
We focus on agents which employ either so called memoryless policies, at which they choose their actions based solely on their current observation \citep{SinghEtAl1994}, or they use a short and fixed history of current and past observations and actions to base the current action on. This has the advantage of being simple to act on \citep{WilliamsSingh1999} and they are easy to realize at no or little additional computational cost.

To highlight the broad applicability of our dynamics we study the emerging learning behavior across five partially observable environment classes.
We find a variety of effects caused by partial observability which generally depend on the environment and its representation. For instance, partial observability  can lead to better learning outcomes faster in a single-agent renewable resource harvesting task, stabilize a chaotic learning process in a multistate zero-sum game and even overcome social dilemmas.
Compared to fully observant agents, partially observant learning often requires more exploration and less weight on future rewards to obtain the favorable learning outcomes.
Furthermore, our method allows the application of dynamical systems theory to partially observable multiagent leaning. We find that partial observability can cause the emergence of catastrophic limit cycles, a critical slowing down of the learning processes between reward regimes and the separation of the learning dynamics into fast and slow eigendirections.
We hope that the presented dynamics become a practical, lightweight and robust tool to systematically investigate the effect of uncertainty of interacting agents.

\section{Background}
\label{sec:Background}

\subsection{Partially observable stochastic games}

\paragraph{Definition.}
The game $G=\langle N, \mathcal S, \tens{\mathcal A}, \tens{\mathcal O}, T, \tens R, \tens O \rangle$ is a stochastic game with $N \in \mathbb N$ agents.
The environment consists of $Z \in \mathbb N$ states $\mathcal S=(S_1, \dots, S_Z)$.
In each state $s$, each agent $i \in \{1,\dots, N\}$ has $M \in \mathbb N$ available actions $\mathcal A^i = (A^i_1,\dots,A^i_M)$ to choose from. $\tens{\mathcal A} = \bigotimes_i \mathcal A^i$ is the joint-action set and agents choose their actions simultaneously. A joint action is denoted by $\tens a = (a^1,\dots,a^N) \in \tens{\mathcal A} $.
With $\tens a^{-i} = (a^1,\dots,$ $a^{i-1},$ $a^{i+1},$ $\dots,$ $a^N)$ we denote the joint action except agent $i$'s.
We chose an identical number of actions for all states and all agents out of notational convenience.
Throughout this paper, we restrict ourselves to ergodic environments without absorbing states.

The transition function $T: \mathcal S \times \tens{\mathcal A} \times \mathcal S \rightarrow [0, 1]$
determines the probabilistic state changes.
$T(s, \tens a, s')$ is the transition probability from current state $s$ to next state $s'$ under joint action $\tens a$.

The reward function $\tens R: \mathcal S \times \tens{\mathcal A} \times \mathcal S \rightarrow \mathbb{R}^N$ maps the triple of current state $s$, joint action $\tens a$ and next state $s'$ to an immediate reward scalar for each agent. $R^i(s,\tens a,s')$ is the reward agent $i$ receives.

Instead of observing the states $s \in \mathcal S$ directly, each agent $i$ observes one of $Q \in \mathbb N$ observations $\mathcal O^i = (O^i_1, \dots, O^i_Q )$ according to the observation functions $O^i: \mathcal S \times \mathcal O^i \rightarrow [0, 1]$. $O^i(s, o)$ is the probability that agent $i$ observes observation $o \in \mathcal O^i$ given that the environment is in state $s \in \mathcal S$.
$\tens{\mathcal O} = \bigotimes_i \mathcal O^i$ is the joint observation set and  $\tens  O = \bigotimes_i  O^i : \mathcal S \times \tens{\mathcal O} \rightarrow [0, 1]^N$ is the joint observation function.
We chose an identical number of observations for all agents out of notational convenience.
By construction, this observation function can model both noisy state observations ($Q=Z$) and hidden states ($Q<Z$).
%

\paragraph{Policies.} We consider agents that choose their actions probabilistically according to their memoryless policy $X^i :
\mathcal O^i \times \mathcal A^i \rightarrow  [0,1]$.
$X^i(o^i, a^i)$ is the probability that agent $i$ chooses action $a^i$ given that it observed observation $o^i$. We denote the joint policy by $\tens X = \tens X(\tens o, \tens a) = \bigotimes_i X^i(o^i, a^i) : \tens{\mathcal O} \times \tens{\mathcal A} \rightarrow [0,1]^N$.


\paragraph{Histories.}
Besides memoryless policies we also consider policies with fixed histories $\mathcal H_\mathsf{h}$ of type $\mathsf h$. The type $\mathsf h$ is composed of $\mathsf h = \mathsf{h}_o \times \mathsf{h}_{\tens a} $ with $\mathsf{h}_o \in \mathbb{N}$ and $\mathsf{h}_{\tens a} \in \mathbb{N}^N$. $\mathsf{h}_o$ represents how many of current and past observations are to be used to encode the histories. Likewise,  $\mathsf{h}_{\tens a}$ represents how many past actions of each agent are to be encoded in the histories. For example, the default memoryless policy is of type $\mathsf h = (1, \tens 0)$. Practically, histories induce an embedding of the game into a larger state space at which the histories $\mathcal H_\mathsf{h}$ correspond to the larger state set and transitions, rewards and observations are adjusted accordingly.

\subsection{Temporal-difference reinforcement learning}
Temporal-difference $Q$-learning is one of the most widely studied reinforcement learning processes \citep{WatkinsDayan1992,schultz1997neural,SuttonBarto2018}. Agents successively improve their evaluations of the quality of the available actions.
Originally developed under the assumption that agents can observe the true Markov state of the environment, we here present the basic temporal-difference $Q$-learning algorithm in the more general formulation, where agents use observations instead of states. When observations exactly map onto the states, the original algorithm is recovered.

At time step $t$ agent $i$ evaluates action $a^i$ at observation $o^i$ to be of quality $Q^i_t(o^i, a^i)$. Those state-action values $Q^i_t(o^i, a^i)$ are then updated after selecting action $a^i_t$ after observing observation $o^i_t$ according to
\begin{equation}
	Q^i_{t+1}(o^i_t,a^i_t) = Q^i_{t}(o^i_t,a^i_t) + \alpha \cdot \delta^i_t(o^i_t,a^i_t),
	\label{eq:QUpdate}
\end{equation}
with the temporal-difference error
\begin{equation}
	\delta^i_t(o^i_t,a^i_t) :=  (1-\gamma)r^i_t +  \gamma\max_b Q^i_t(o^i_{t+1}, b) - Q^i_t(o^i_t,a^i_t).
	\label{eq:TDe}
\end{equation}

The \textit{discount factor} parameter $\gamma \in [0,1) $ regulates how much the agent cares for future rewards.
The \textit{learning rate} parameter $\alpha \in (0,1)$ regulates how much new information is used for an observation-action-value update.
For the sake of simplicity, we assume identical parameters across agents throughout this paper and therefore do not equip parameters with agent indices.
The variable $r^i_t$ refers to the immediate reward at time step $t$.
Note that the $(1-\gamma)$ prefactor in front of the reward occurs when we assume that agents aim to maximize a return defined as $G^i_t = (1-\gamma) \sum_{k=0}^\infty \gamma^k r^i_{t+k}$ \citep{BarfussEtAl2019}. This leads the values to be on the same scale as the rewards.

Agents select actions based on the current observation-action values $Q^i_t(o^i, a^i)$ balancing exploitation (i.e., selecting the action of maximum quality) and exploration (i.e., selecting lower quality actions in order to learn more about the environment). We here use the widely used Boltzmann policy function. The probability of choosing action $a^i$ under observation $o^i$ is
\begin{equation}
	X^i_t(o^i, a^i) = \frac{e^{\beta Q^i_t(o^i, a^i)}}{\sum_{b \in \mathcal A^i} e^{\beta Q^i_t(o^i, b)} },
	\label{eq:X}
\end{equation}
where the \textit{intensity of choice} parameter $\beta$ controls the exploration-exploitation trade-off.
%
Throughout this paper, we are interested in the idealized learning process with fixed parameters $\alpha$, $\beta$ and $\gamma$ throughout learning and evaluating a policy.

\section{Derivation}
\label{sec:Derivation}


In this section we derive the deterministic reinforcement learning dynamics under partial observability in discrete time.
As classic evolutionary dynamics operate in the theoretical limit of an infinite population, the learning dynamics are derived by considering an infinite memory batch \citep{barfuss2020reinforcement,barfuss2021dynamical}.
A learning dynamic update of the current policy uses policy-averages instead of individual samples. 
%
Thus, we need to construct the policy-average temporal-difference error $\bar\delta^i$ to be inserted in the
update for the joint policy,
\begin{equation}
	X^i_{t+1}(o^i, a^i) = \frac{X^i_t(o^i, a^i) \cdot \exp[\alpha\beta \bar\delta^i(o^i, a^i)]}
	{\sum_b X^i_t(o^i,b) \cdot \exp[\alpha\beta \bar\delta^i(o^i, b)] }.
	\label{eq:XUpdate}
\end{equation}
Equation~\ref{eq:XUpdate} can be derived by combining Eqs. \ref{eq:QUpdate} and \ref{eq:X}. The bar on top of $\delta^i$ indicates implicitly that $\bar\delta^i$ depends fully on the current joint policy $\tens{X}_t$. Computing $\bar\delta^i(o^i, a^i)$ involves averaging over policies, environmental transitions and observations for the first two terms of the temporal-difference error (Eq. \ref{eq:TDe}), the immediate rewards and the qualities of the next observation.
The quality of the current observation, $Q^i_t(o^i_t,a^i_t)$ becomes $\beta^{-1} \ln X^i(o^i, a^i)$ in the average temporal-difference error and serves as regularization term. This can be derived by inverting Eq. \ref{eq:X} and realizing that the dynamics induced by Eq. \ref{eq:XUpdate} are invariant under additive transformations which are constant in actions. 

\subsection{Beliefs}
The challenge is that the rewards $R^i(s,\tens a, s')$ in the stochastic game model depend on the true states, not on the observations of the agents. Thus, in order to obtain the average observation-action rewards $\bar{R}^{i}(o^i, a^i)$, we need a mapping from observations to states. The observation function is a mapping from states to observations. With Bayes's rule,
\begin{equation}
	\bar B^i(o^i, s) = \frac{O^i(s, o^i) \bar P(s)}{\sum_s O^i(s, o^i) \bar P(s)},
	\label{eq:Bios}
\end{equation}
we can transform the observation function into a belief function, following the rules of probability. $\bar B^i(o^i, s)$ is the belief of agent $i$ (or simply the probability) that the environment is in state $s$ when it observed observation $o^i$.

The only problem is how to obtain the policy-average stationary state distribution $\bar P(s)$. $\bar P(s)$ is the left-eigenvector of the average transition matrix $\bar T(\tens s, \tens s)$ where the entry $\bar T(s, s')$ denotes the probability of transitioning from state $s$ to state $s'$.
This matrix could be obtained as $\bar T(s, s') = \prod_j \sum_{a^j} \bar Y^j(s, a^j) T(s, \tens a, s')$ if we had the probability for each agent $j$ to choose action $a^j$ in state $s$, $\bar Y^j(s, a^j)$. However, we assumed that agents condition their actions only on observations, $X^j(o^j, a^j)$. Yet,
whenever the environment is in state $s$, agent $j$ observes observation $o^j$ with probability $O^j(s, o^j)$ and than chooses action $a^j$ with probability $X^j(o^j, a^j)$. Thus, with
\begin{equation}
	\bar Y^j(s, a^j) :=
	\sum_{o^j \in \mathcal O^j} O^j(s, o^j) X^j(o^j, a^j),
	\label{eq:Yisa}
\end{equation}
we can average out the observation and obtain the policy-average state policies $\bar Y^j(s, a^j)$. Note that $\bar Y^j(s, a^j)$ are proper conditional probabilities, which can be seen by applying $\sum_{a^j}$ to both sides of Eq. \ref{eq:Yisa}. With $\bar Y^j(s, a^j)$ we can then compute the policy-average transition matrix  $\bar T(\tens s, \tens s)$, its left-eigenvector, the stationary state distribution $\bar P(s)$, and thus, the policy-average belief of agent $i$ that the environment is in state $s$ when it observed observation $o^i$, $\bar B^i(o^i, s)$.

\subsection{Rewards}
Whenever agent $i$ observes observation $o^i$, with probability $\bar B^i(o^i,s)$ the environment is in state $s$ where all other agents $j\neq i$ behave according to $\bar Y^j(s, a^j)$, the environment transitions to a next state $s'$ with probability $ T(s, \underline a, s')$, and agent $i$ receives the reward $R^i(s, \underline a, s')$. Mathematically, the policy-average reward for action $a^i$ under observation $o^i$ reads
\begin{align}
	\bar{R}^{i}(o^i, a^i) :=   \sum_s \sum_{a^j} \sum_{s'} \prod_{j \neq i}
	\bar B^i(o^i, s) \bar Y^j(s, a^j) \nonumber \\ T(s, \underline a, s') R^i(s, \underline a, s').
\end{align}


\subsection{Qualities}
Second, the policy-average of the quality of the next observation ($\max_b Q^i_t(o^i_{t+1}, b)$ in Eq.~\ref{eq:TDe}) is computed by averaging over all states, all actions of the other agents, next states and next observations. Whenever agent $i$ observers observation $o^i$,  the environment is in state $s$ with probability $\bar B^i(o^i, s)$. There, all other agents $j \neq i$ choose their action $a^j$ with probability $\bar Y^j(s, a^j)$. Consequently, the environment transitions to the next state $s'$ with probability $T(s, \tens a, s')$. At $s'$, the agent observes observation $o'$ with probability $O^i(s', o')$ and estimates the quality to be of value $\max_b \bar Q^i(o', b)$. Mathematically, we write
\begin{align}
	{}^{\text{max}}\!\bar{Q}^i(o^i, a^i) := \sum_s \sum_{a^j} \sum_{s'} \sum_{o'} \prod_{j \neq i}
	\bar B^i(o^i, s) \bar Y^j(s, a^j) \nonumber \\
	T(s, \underline a, s') O^i(s', o') \max_b \bar Q^i(o',b).
\end{align}

Here, we replace the quality estimates $Q^i_t(o^i, a^i)$, which evolve in time $t$ (Eq.~\ref{eq:QUpdate}), with the policy-average observation-action quality $\bar Q^i(o^i, a^i)$, which is the expected discounted sum of future rewards from executing action $a^i$ at observation $o^i$ and then following along the joint policy $\underline X$.
It is obtained by a discount factor weighted average of the current policy-average reward $\bar R^i(o^i, a^i)$ and the policy-average observation quality of the next observation $\bar V^i(o')$,
\begin{align}
	\bar Q^i(o^i, a^i) =& \ (1-\gamma) \bar R^i(o^i, a^i) \nonumber \\
	&+ \ \gamma \sum_{o' \in \mathcal O^i}
	\bar T^i(o^i, a^i, o' ) \bar V^i(o').
	\label{eq:Qioa}
\end{align}
Here, $\bar T^i(o^i, a^i, o' )$ is agent $i$'s policy-average transition probability of observing observation $o'$ at the next time step given it observed observation $o^i$ at the current time step and chose action $a^i$.
It is computed by averaging over all states, next states and all actions of the other agents. Whenever agent $i$ observes observation $o^i$ and selects action $a^i$, the environment is in state $s$ with probability $\bar B^i(o^i, s)$, where all other agents $j \neq i$ select action $a^j$ with probability $\bar Y^j(s, a^j)$. Consequently, the environment will transition to the next state $s'$ with probability $T(s, \tens a, s')$ which is observed with probability $O^i(s', o')$ as $o'$ by agent $i$. Mathematically, we write
\begin{align}
	\bar T^i(o^i, a^i, o' ) = \sum_s  \sum_{a^j} \sum_{s'} \prod_{j \neq i} \bar B^i(o^i, s) \bar Y^j(s, a^j) \nonumber \\
	T(s, \underline a, s') O^i(s', o').
\end{align}

\begin{figure*}
	\vskip 0.2in
	\begin{center}
		\centerline
		{\vcenteredhbox{
				\includegraphics[draft=False, 	width=0.2\linewidth]{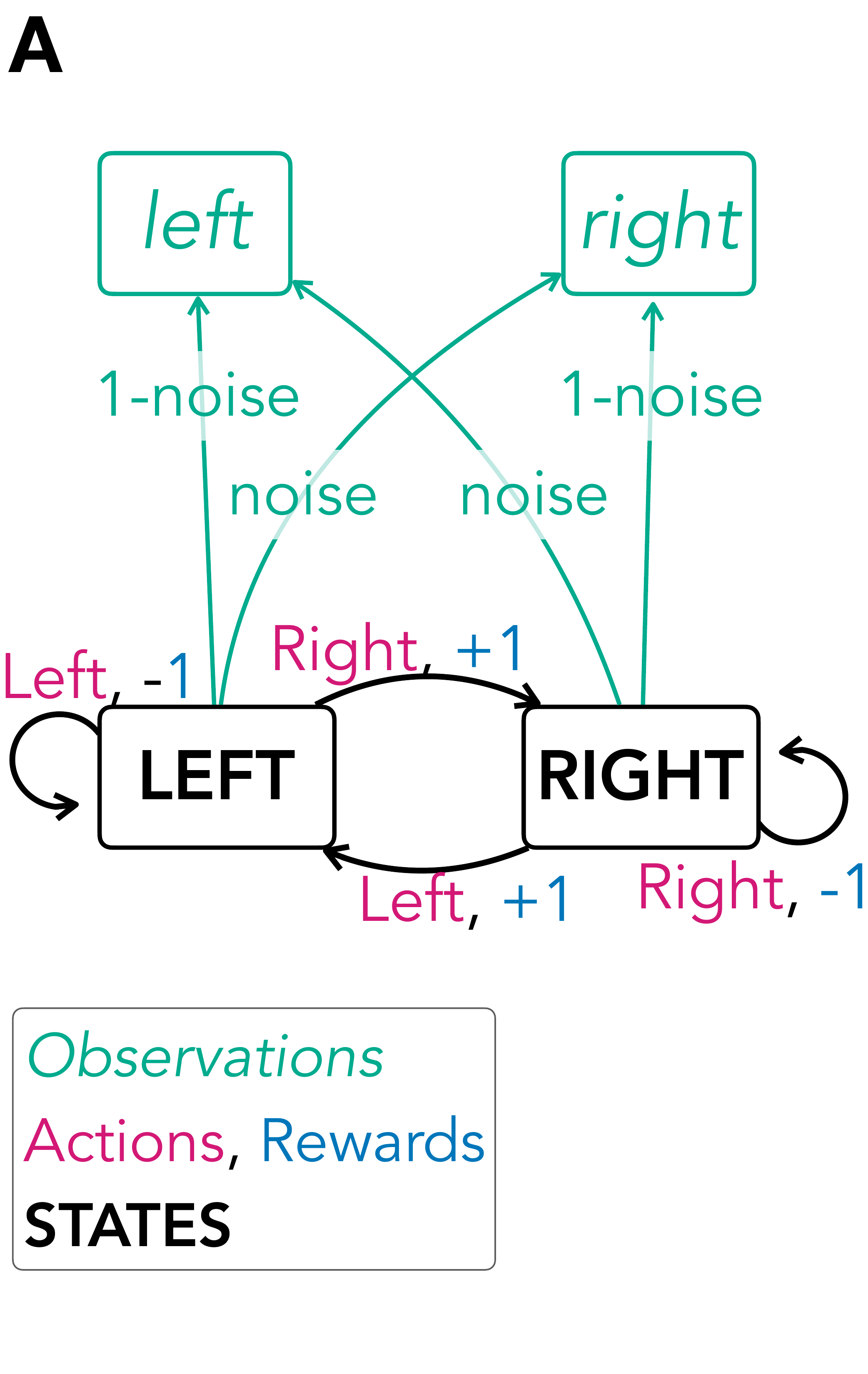}}
			\hspace{0.18cm}
			\vcenteredhbox{
				\includegraphics[draft=False, width=0.75\linewidth]{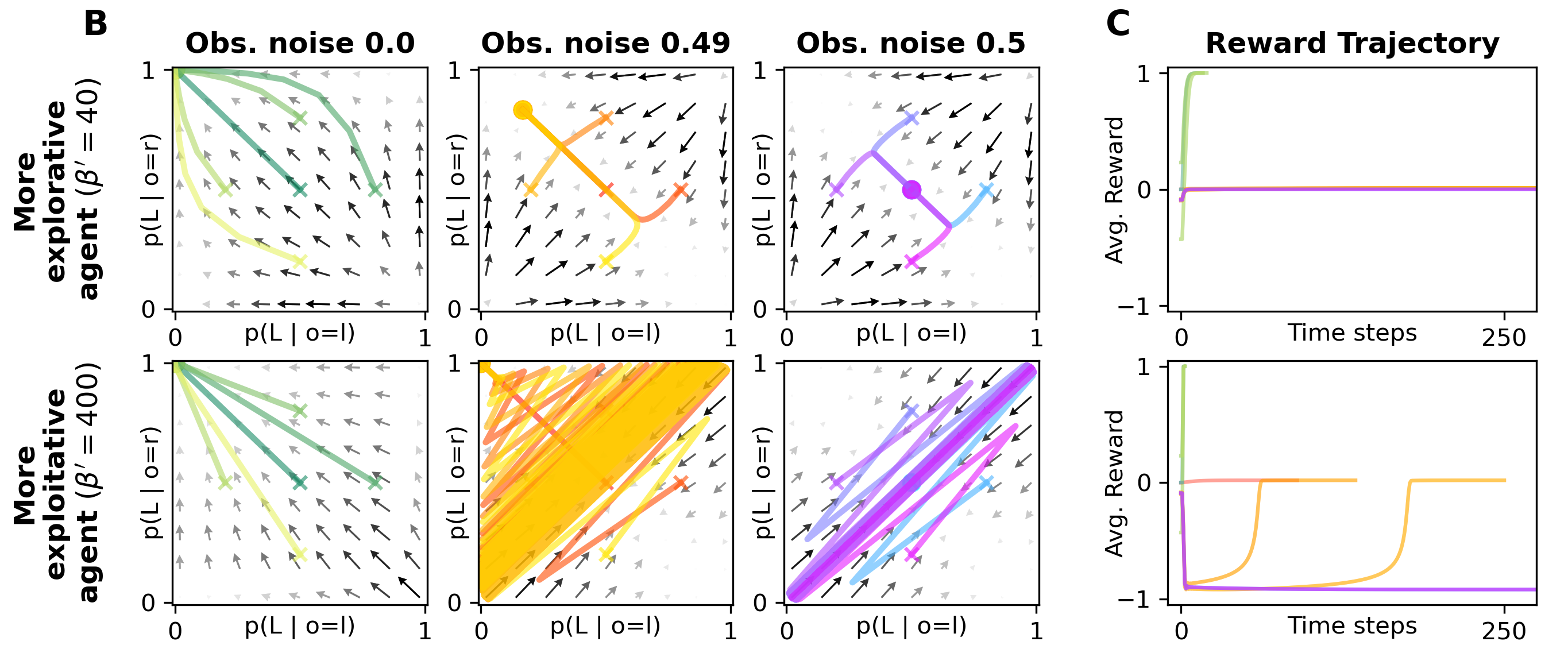}}}
			\vspace{-0.25cm}
		\caption{\textbf{Deterministic learning dynamics in a simple coordination task.} The agent has to anti-coordinate its actions on the environmental state, which it observes through a noisy channel (A).  Learning trajectories are computed for three observational noise levels (0.0, 0.49, 0.5) and two intensities of choice, a low one ($\beta'=\beta/(1-\gamma)=40$), shown in the top row, and a high one ($\beta'=400$), shown in the bottom row.
		Panel B shows the corresponding policy spaces in which the agent's probability of choosing $\textsf{Left}$, given the agent perceived the environment to be in the \textit{left} state, is plotted on the $x$-axes; and the agent's probability of choosing $\textsf{Left}$, given the agent perceived the environment to be in the \textit{right} state, is plotted on the $y$-axes. 5 individual trajectories, whose initial policies were centered around the center of the policy space, are plotted in color. Arrows in gray indicate the flow of the learning dynamical system.
		Panel C shows the corresponding reward trajectories. Remaining hyperparameters were set as $\alpha=0.01$, $\gamma=0.9$.
		Partial observability can cause the learning to enter low-rewarding limit cycles under high intensity of choice.
		}
		\label{fig:SimpleCoordination}
	\end{center}
	\vskip -0.2in
\end{figure*}

Further at Eq. \ref{eq:Qioa}, $\bar V^i(o^i)$ is the policy-average observation quality, i.e., the expected discounted sum of future rewards from observation $o^i$ and then following along the joint policy $\underline X$. They are computed via matrix inversion according to
\begin{equation}
	\bar V^i(\tens o) = (1-\gamma) [
	\tens{\mathds{1}}_Q - \gamma  \bar T^i(\tens{o}, \tens{o}) ]^{-1} \bar R(\tens{o}).
	\label{eq:Vio}
\end{equation}
This equation is a direct conversion of the Bellman equation
$\bar V^i(o^i) = (1-\gamma) \bar R(o^i) + \gamma \sum_{o'} \bar T^i(o^i, o' ) \bar V^i(o') $, which expresses that the value of the current observation is the discount factor weighted average of the current reward and the value of the next observation. Underlined observation variables indicate that the corresponding object is a vector or matrix and $\underline{\mathds{1}}_Q$ is a $Q$-by-$Q$ identity matrix.

$\bar T^i(\underline o, \underline o)$ denotes the policy-averaged transition matrix for agent $i$. The entry $\bar T^i(o^i, o')$ indicates the probability that agent $i$ will observe observation $o'$ after observing observation $o^i$ at the previous time step, given all agents follow the joint policy $\tens X$. We compute them by averaging over all states, all actions from all agents and all next states,
\begin{align}
	\bar T^i(o^i, o') = \sum_s \sum_{a^j} \sum_{s'} \prod_{j} \bar B^i(o^i, s) \bar Y^j(s, a^j) \nonumber \\
	T(s, \underline a, s') O^i(s', o').
	\label{eq:Tioo}
\end{align}
For any observation $o^i$, $\bar B^i(o^i, s)$ is the probability to be in state $s$, where all agents $j$ act according to $\bar Y^j(s, a^j)$. Therefore, the environment transitions with probability $T(s, \underline a, s')$ from state $s$ to the next state $s'$, which is observed by agent $i$ as observation $o'$ with probability $O^i(s', o')$. Note that $\bar T^i(\tens o, \tens o)$ is a proper probabilistic matrix. This can be seen by applying $\sum_{o'}$ to both sides of Eq. \ref{eq:Tioo}.

Further in Eq. \ref{eq:Vio}, $\bar{R}^i(o^i)$ denotes the policy-average reward agent $i$ obtains from observation $o^i$. We compute them by averaging over all states, all actions from all agents and all next states.  Whenever agent $i$ observes observation $o^i$, the environment is in state $s$ with probability $\bar B^i(o^i, s)$. Here, all agents $j$ choose action $a^j$ with probability $\bar Y^j(s, a^j)$. Hence, the environment transitions to the next state $s'$ with probability $T(s, \tens a, s')$ and agent $i$ receives the reward $R^i(s, \tens a, s')$,
\begin{align}
	\bar{R}^{i}(o^i) := \sum_s \sum_{a^j} \sum_{s'} \prod_{j}
	\bar B^i(o^i, s) \bar Y^j(s, a^j) \nonumber \\ T(s, \underline a, s') R^i(s, \underline a, s').
\end{align}

Note that the quality ${}^\text{max}\!\bar{Q}^i(o^i, a^i)$ depends on $o^i$ and $a^i$ although it is the policy-averaged maximum observation-action value of the next observation.

\subsection{Temporal-difference error}
All together, the policy-average temporal-difference error, to be inserted into Eq. \ref{eq:XUpdate}, reads
\begin{align}
	\bar\delta^i(o^i, a^i) = (1-\gamma) \bar R^i(o^i, a^i) + \gamma {}^{\text{max}}\!\bar{Q}^i(o^i, a^i) \nonumber \\ - \frac{\ln X^i(o^i, a^i)}{\beta}.
	\label{eq:detTDe}
\end{align}

\section{Experiments}
\label{sec:Experiements}
We study the emerging learning dynamics across five test environments: three single-agent decision problems and two multiagent games. Three environments will cover noisy observations, the other two focus on a reduced observation space, where a given observation is consistent with multiple true states of the world.
As one evaluation metric we use the average reward, $\sum_s \bar P(s) \bar R^i(s)$, where $\bar P(s)$ is the stationary state-distribution and $\bar R^i(s) = \sum_{a^j} \sum_{s'} \prod_j \bar Y^j(s, a^j) T(s, \tens{a}, s') R^i(s, \tens{a}, s')$ is the average reward for each state given the current policy $\tens X$ (see Sec. \ref{sec:Derivation}).
We defined a learning trajectory as having converged if the norm between old and updated policy (according to Eq.~\ref{eq:XUpdate}) is below $10^{-5}$.
Since we defined the return with the $(1-\gamma)$ prefactor we also consider a scaled version of the intensity of choice parameter $\beta=\beta' / (1-\gamma)$ for some experiments. Doing so preserves the ratio of exploration and exploitation in the temporal-difference error (Eq. \ref{eq:detTDe}) under changes in the discount factor $\gamma$.

\subsection{Simple coordination task}
\label{sec:Coordination}

\paragraph{Environment description.}
The first environment is a simple coordination task in which the agent must move between the left and right environmental state in order to obtain a maximum reward of 1. Coordinating which of the two available actions ($\mathsf{Left, Right}$) to choose from is complicated by observational noise $\nu$, letting the agent perceive the correct state only with probability $1-\nu$ (Fig.~\ref{fig:SimpleCoordination}\,A). This environment is adapted from \citet{SinghEtAl1994}.

\paragraph{Results.} Figure~\ref{fig:SimpleCoordination} shows how partial observability can cause the deterministic learning dynamics to enter low-rewarding limit cycles under a high intensity of choice.
Often, learning a policy involves a trade-off between the amount of reward from that policy and the amount of time required learning it. In the simple coordination task with  perfect observation, a high intensity of choice
can speed up the learning process by a factor of 6. The trajectories with $\beta'=40$ require about 18 time steps to arrive at the optimal policy with average reward 1 (green lines, top row), the trajectories with $\beta'=400$ require only 3 time steps (green lines, bottom row).
Thus, a high intensity of choice is clearly preferable under perfect observation.

With fully uninformative observations (observational noise level $\nu=0.5$, Fig.~\ref{fig:SimpleCoordination}\,B, third column) a more explorative agent (i.e., lower intensity of choice, top row) has an advantage. From all initial policies, it takes the agent about 580 time steps to learn to fully randomize its actions. This yields an average reward of zero and is also the optimal memoryless policy \citep{SinghEtAl1994}.
The more exploitative agent (bottom row) on the other hand enters a limit cycle between choosing $\mathsf{Left}$ and $\mathsf{Right}$ almost deterministically, irrespective of its current observations. Thus, while choosing $\mathsf{Left}$ the agent is trapped in the LEFT state, obtaining an average reward of $\sim\!-1$. While choosing $\mathsf{Right}$, the agent is trapped in the RIGHT state also obtaining an average reward of $\sim\!-1$. The positive reward obtained through the move between states is neglected, since the derived dynamics consider the theoretical limit of an an infinite memory batch \citep{barfuss2020reinforcement}. This can also be interpreted as a complete separation of the interaction timescale and the adaptation timescale \citep{BarfussEtAl2019,barfuss2021dynamical}. The agent experiences an infinite amount of negative reward during interaction and only one single positive reward after the policy adaptation. It will be interesting to reexamine this scenario under relaxed conditions when interaction and adaption timescales are not completely separated in future work.

When observations are almost completely noisy, yet still contain some information about the true environmental state, the more exploitative agent learns a slightly more rewarding policy faster
(Fig.~\ref{fig:SimpleCoordination}\,B second column).
An observational noise level of $\nu = 0.49$ means that of 100 times being in the LEFT environmental state, the agent will observe on average \textit{left} 51 times and \textit{right} 49 times. Here, from all initial policies, the more explorative agent (top row) converges to a fixed point in the upper left part of the policy space in about 600 time steps, i.e., slower than under completely noisy observations. This policy yields an average reward of about 0.013. The more exploitative agent (bottom row) learns on an interesting transient resembling the limit cycle of the fully uninformative case, yet manages to converge to the deterministic policy in the upper left of the policy space. This yields an average reward of about 0.02 and takes at most 250 times, depending on the initial policy. This is still distinctly faster than the more explorative agent.

Overall, it is interesting to observe how partial observability caused a well-known dynamical-systems phenomenon in the learning dynamics, which can explain the abrupt improvements in the reward trajectories (Fig.~\ref{fig:SimpleCoordination}\,C bottom): the separation of the dynamics into a fast eigendirection along the diagonal from the bottom left to the top right of the policy space and a slow eigendirection perpendicular to that \citep{Strogatz2018}. The slow eigendirection corresponds to a coordinated policy where the agent's observation is decisive for its actions. Along the fast eigendirection the agent's policy is independent of its observations. The more explorative agent moves along these axes whereas the more exploitative agent overrides. Yet, as long as there is some information in the observations about the environmental state, the more exploitative agent learns better policies faster.

\begin{figure}
	\begin{center}
		\includegraphics[draft=False, width=0.8\linewidth]{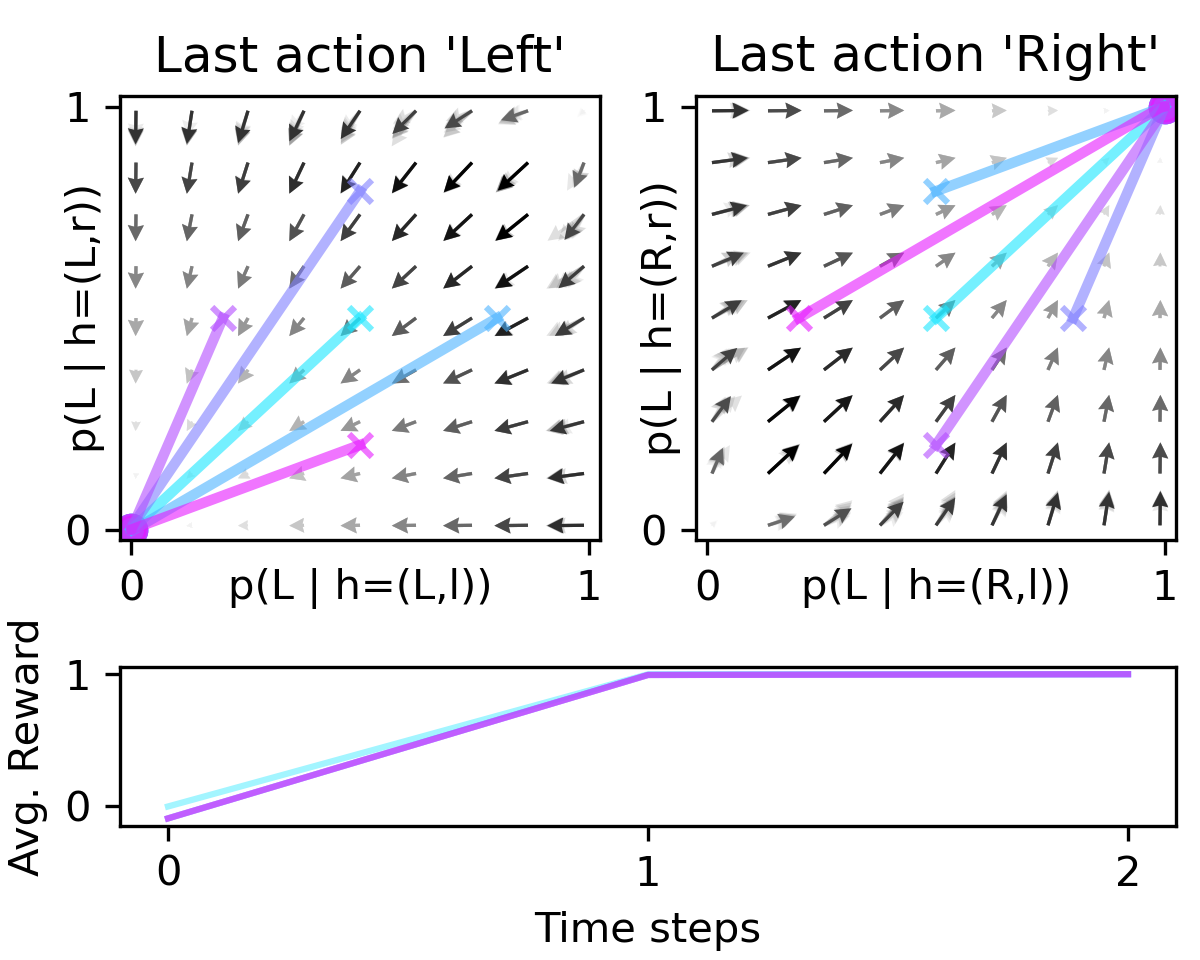}
		\caption{\textbf{Deterministic learning dynamics with history in a simple coordination task.} Same situation as in Fig.~\ref{fig:SimpleCoordination} with high intensity of choice $\beta'=400$ and observational noise level $0.5$ but here, the agent remembers and conditions its action not only on the last observation but also on its last action. Thus, the left (right) panel shows
			a projection of the learning dynamics, given the last action was $\textsf{Left}$ ($\textsf{Right}$). The $x$-axes show the probabilities of choosing $\textsf{Left}$, given the last environmental observation was \textit{left}. The $y$-axes show the probabilities of choosing $\textsf{Left}$, given the last environmental observation was \textit{right}. The agent learns to alternate between $\textsf{Left}$ and $\textsf{Right}$, which yield the highest reward possible, in (at most) only two time steps.}
		\label{fig:SimpleCoordinationHistory}
	\end{center}
\end{figure}

So far we examined only memoryless policies, i.e., policies that condition their choice of action only on the current observation. If the more exploitative agent ($\beta' = 400$) is able to condition its choice of action not only on the current observations but also on its last action, then it learns the optimal policy with an average reward of 1 in at most only two time steps - even under fully uninformative state observations (Fig. \ref{fig:SimpleCoordinationHistory}). The agent learns to alternate between $\textsf{Left}$ and $\textsf{Right}$. This learned policy and even the whole learning dynamics do not depend on the state-observation, as shown by the straight line trajectories and corresponding dynamical flow arrows in Fig. \ref{fig:SimpleCoordinationHistory}.

\begin{figure}
	\begin{center}
		\includegraphics[draft=False, width=0.99\linewidth]{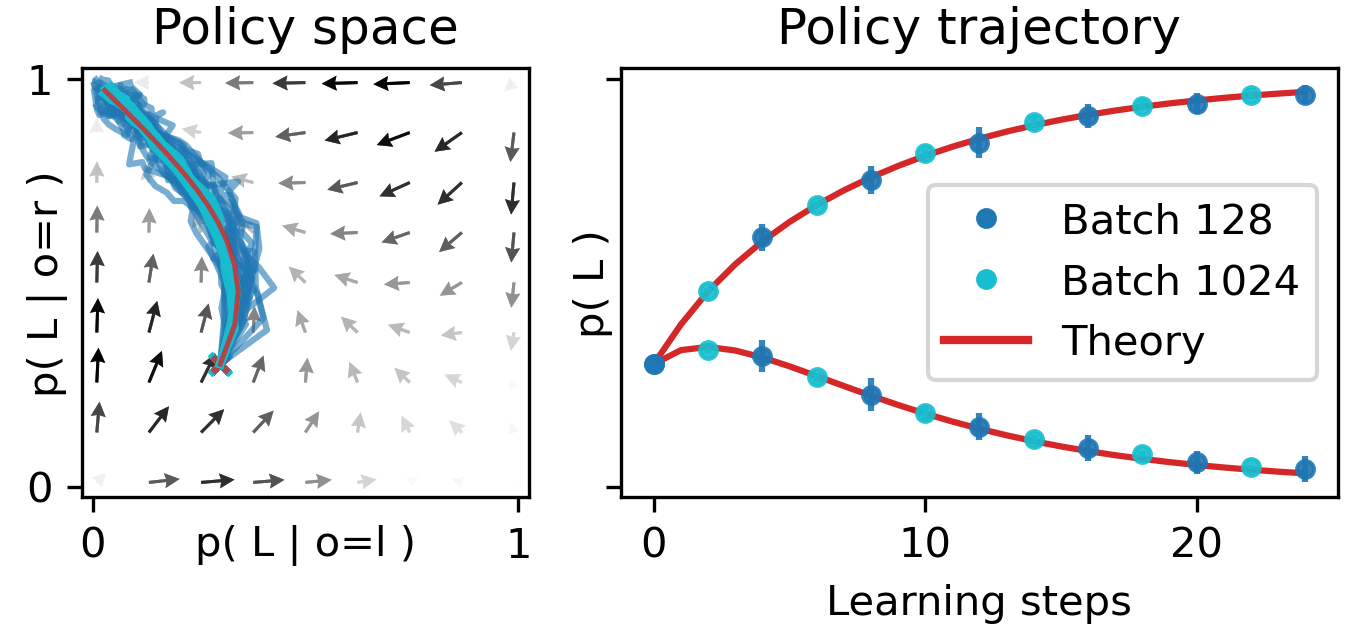}
		\caption{\textbf{Comparison between deterministic learning dynamics and batch-learning algorithm} in a simple coordination task with noise level 0.4. The learning trajectory from an initial policy of choosing $\mathsf{Left}$ with $30\%$ given both observations is shown for a batch reinforcement learning algorithm with batch sizes 128 (blue; sample size 25) and 1024 (cyan; sample size 5) as well as for the deterministic theory (red); in policy space (left) and versus learning steps (right). On the right only every forth learning step of the batch learners mean policy values are plotted. Error bars indicate the standard deviation. Hyperparameters are $\alpha=0.01, \beta'=40,$ and $\gamma=0.09$. The deterministic learning dynamics are well approximated by the batch-learning algorithm.}
		\label{fig:SimpleCoordinationBatch}
	\end{center}

\end{figure}

As a consistency check we compare the derived deterministic learning dynamics with partial observability to a sample-batch reinforcement learning algorithm, as detailed in Ref.~\cite{barfuss2021dynamical} (Fig.~\ref{fig:SimpleCoordinationBatch}).
The batch learning algorithm collects observation and reward experiences inside a batch of size $K$ while keeping its policy fixed before it then updates its policy using the whole information of the collected batch.
This is a widely occurring principle for improved data efficiency and learning stability  \cite{lange2012batch} and is used for example in memory-replay \cite{lin1992self} and model-based reinforcement learning \cite{sutton1990integrated}. Figure~\ref{fig:SimpleCoordinationBatch} shows that our deterministic theory describes such batch learning approaches well under large batch sizes.
Yet the calculation time of the deterministic dynamics was in the order of 100 times faster than the simulation of the algorithms.


\begin{figure*}
	\vskip 0.2in
	\begin{center}
		\includegraphics[draft=False, width=0.99\linewidth]{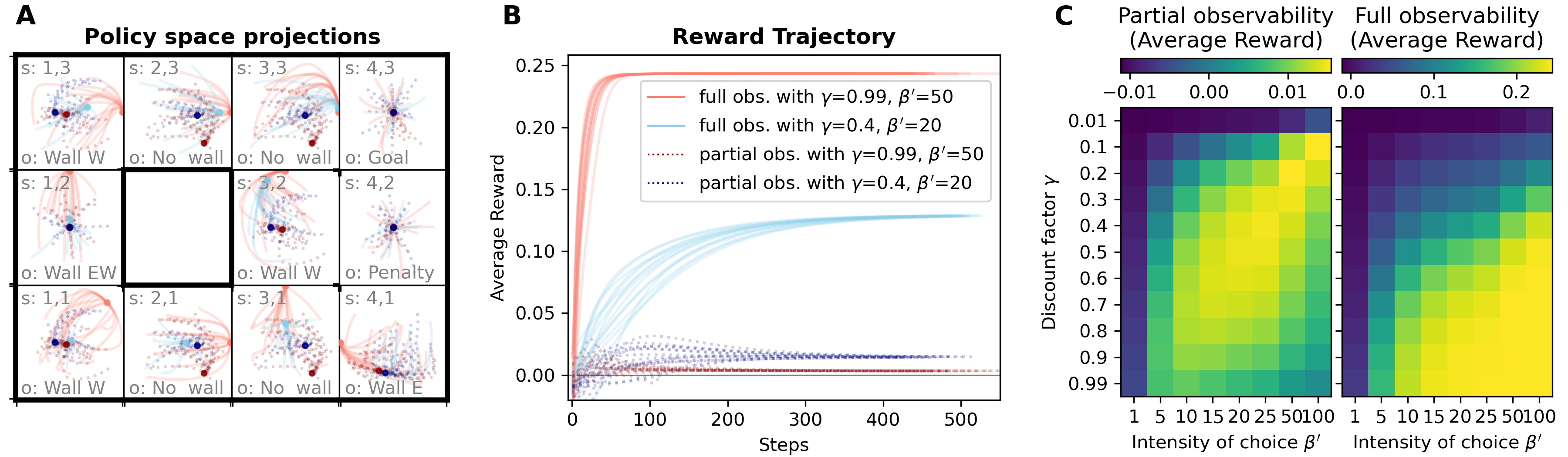}
		\caption{\textbf{Deterministic learning dynamics in a navigation task.}
			Learning trajectories under both partial (dashed lines) and full observability (straight lines) are plotted for various hyperparameter combinations from 15 random initial policies each. Panel A shows the grid world. The trajectories of the policies' action probabilities are projected into each observation/state, such that a deterministic policy toward one direction appears at the edge of that direction in the center. Stochastic policies appear inside the patches. Panel B shows the corresponding reward trajectories. In Panel C, two hyperparameter grids show the average reward at convergence for an agent with partial and full observability (with independent color scales for each case). The learning rate was set to $\alpha=0.01$. In contrast to a fully observant agent, neither a high weight on future rewards (large $\gamma$) nor a high intensity of choice (large $\beta'$) leads to the highest reward for a partially observant agent. Instead, the highest reward depends on the mutual combination of the two hyperparameters.}
		\label{fig:NavigationTask}
	\end{center}
	\vskip -0.2in
\end{figure*}

\subsection{Navigation task}

\paragraph{Environment description.} The next environment is the single-agent navigation task adapted from \citeauthor{ParrRussel1995}'s Grid World (\citeyear{ParrRussel1995}). It consists of 11 states, 6 observations, 4 actions and 1 agent (Fig.~\ref{fig:NavigationTask}\,A).
The agent can move north, south, east, and west. If the agent would move into a wall, then it stays on its current patch. The agent wants to reach the patch in the upper right, which is rewarded by a reward of $1$. However, entering the patch below is punished by a reward of $-1$. In both cases, the episode ends and the agent begins a new episode on one randomly chosen patch out of the nine other patches. All other state-action combinations yield zero reward.
We use this environment to compare the effect of various hyperparameter combinations on the learning behavior of an agent with partial observability to an agent with full observability.
Under partial observability the agent can only observe whether or not there is a wall east and west of its current patch. Imagine, for example, a robot equipped only with haptic sensors on its sides or an insect with corresponding antennae. With full observability, the agent can distinguish each grid patch separately.

\paragraph{Results.}  In contrast to a fully observant agent, neither a high weight on future rewards (large $\gamma$) nor a high intensity of choice (large $\beta'$) leads to a large reward for a partially observable agent. Instead, the highest reward depends on the mutual combination of the two hyperparameters.
For a hyperparameter combination of $\gamma=0.99$ and $\beta'=50$ an agent with full observability quickly learns the optimal policy (Fig.~\ref{fig:NavigationTask}\,A\&B, light-red straight lines). Observe also how the light-red straight lines in states (3,2) and (3,3) avoid being close to the penalty state. In contrast, less weight on future rewards (a lower discount factor of $\gamma=0.4$) and more exploration with $\beta'=20$ lead to a lower average reward at convergence (light blue straight lines).
Observe how the convergence points in policy space (light blue dots) are increasingly farther apart from the optimal policy (light red dots) the more steps the grid cell is away from the goal.
However, when we turn to the agent with partial observability, it is the other way around. Here less weight on future rewards and more exploration lead to a better average reward at convergence (dark colored dashed lines).
This result can be explained as follows. In a fully observable Markov decision process there is always an optimal deterministic policy \citep{Puterman2014}. See how the light red straight lines converges to edge of most grid cells, indicating a deterministic action in that direction (Fig.~\ref{fig:NavigationTask}\,A). Yet policies in a partially observable Markov decision process often require stochasticity \citep{SinghEtAl1994}. More exploration directly ensures that, although not in a reward-targeted way.
Less weight on future rewards might be advantageous under partial observability since too much weight on too distant rewards in the future cannot pay off when there is a fundamental uncertainty about which state the agent occupies or even about what the real states of the environment are. When the environment is only partially observable, anticipating too distant rewards does not have to be beneficial.

\begin{figure}
	\begin{center}
		\includegraphics[draft=False, width=0.8\linewidth]{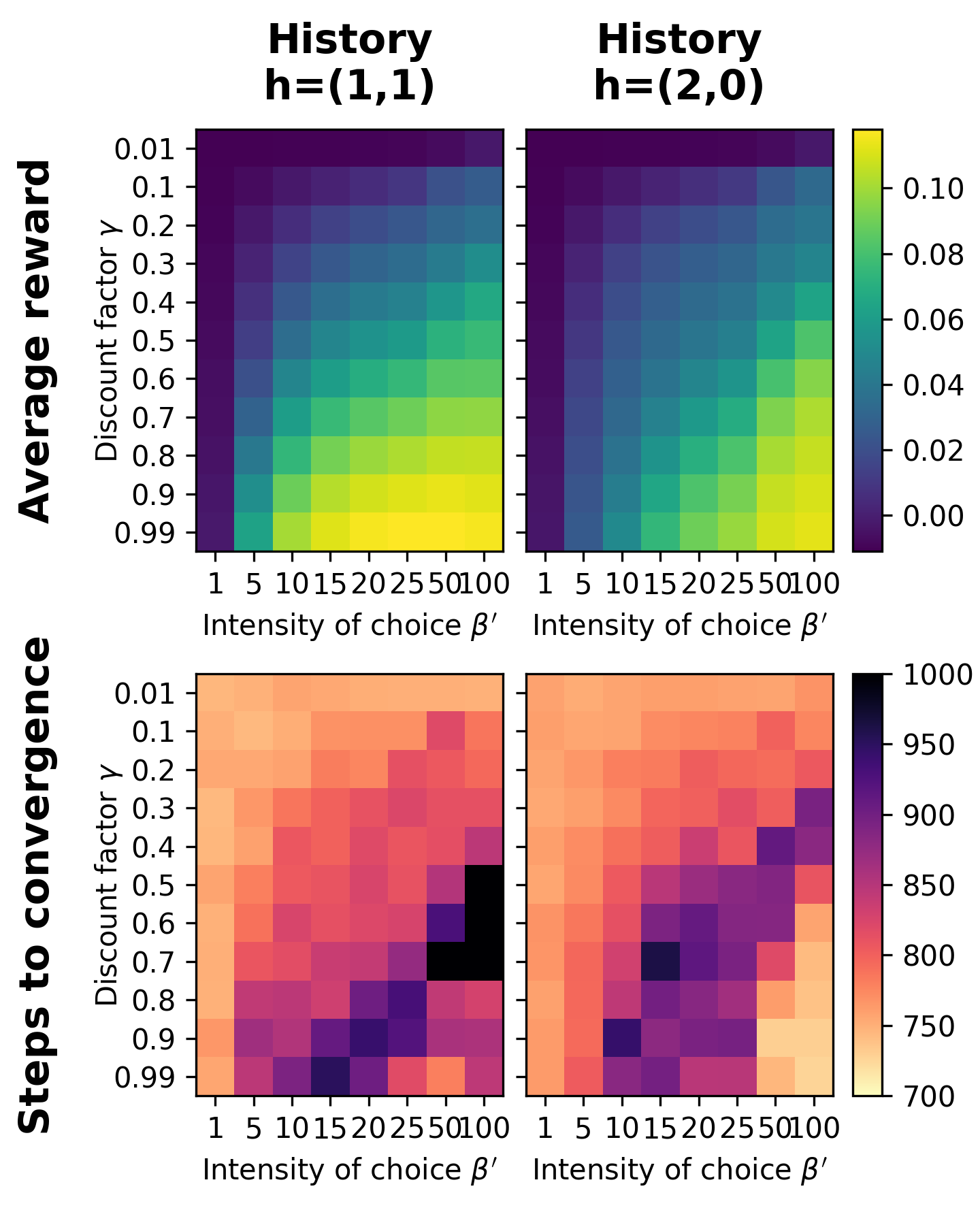}
			\vspace{-0.5cm}
		\caption{\textbf{Hyperparameter grids for the deterministic learning dynamics with history in the navigation task.} Two types of histories are compared. The left plots show results for an agent which conditions its action on the current observation and last action $\mathsf h=(1,1)$. The plots on the right show results for an agent which conditions its action on the current and last observations $\mathsf h=(2,0)$. The top plots show the average reward at convergence, the bottom plots the time steps to convergence, each on the same color scale. Results are averaged over 15 Monte Carlo runs from random initial policies. The learning rate was set to $\alpha=0.01$. Both types of history obtain similar maximum average reward, but at different hyperparameter combinations.}
		\label{fig:NaviHist}
	\end{center}
	\vspace{-1.0cm}
\end{figure}

A systematic analysis of the hyperparameter grid (Fig.~\ref{fig:NavigationTask}\,C) confirms
that partial observability requires the right combination of the two hyperparameters in order to learn the highest reward. Under full observability, simply setting a sufficiently high weight on future rewards $\gamma$ and a sufficiently strong intensity of choice $\beta$ (i.e., little exploration) leads to the average reward of the optimal policy. In contrast, for partial observability too much farsightedness and too intense exploitation can hurt the performance of the agent. Instead, the optimal reward at convergence is obtained by a more randomly explorative and myopic agent.

With memoryless policies, the average reward of the partially observant agent is smaller by an order of magnitude compared to the fully observant agent.
Figure~\ref{fig:NaviHist} compares the results of the two simplest types of history, i.e., where the agent uses one more piece of information. Thus, the agent conditions its actions either on the current observation and the last action $\mathsf h=(1,1)$ or on the current and last observations $\mathsf h=(2,0)$.
Both types of history are able to obtain a similar maximum average reward, with a slight advantage for history $\mathsf h=(1,1)$. Although both are of the simplest type of history conceivable the difference in maximum  reward between the partially observant agent and fully observant agent is already halved, compared to the partially observant agent without history.

Also, the set of hyperparameter combinations that obtain a high average reward is shifted to the lower right corner of the parameter space where also the fully observant agent obtains its maximum. Interestingly, though, the set of high rewarding hyperparameter combinations is not identical across the two types of history. The action-depended history ($\mathsf h=(1,1)$) performs best with a high weight on future rewards $\gamma$ and more exploration, whereas the two-observation history ($\mathsf h=(2,0)$) obtains the highest rewards by more exploitation across a wider range of weights on future rewards $\gamma$.

Furthermore, the learner experiences another dynamical systems phenomenon: a critical slowing down of its learning dynamics \cite{Strogatz2018} before a hyperparameter bifurcation into the high rewarding regime (Fig.~\ref{fig:NaviHist}, bottom row). In the area around the hyperparameter regions which obtain high average reward (yellow area in the plots in the top row), the number of time steps it takes the agent to converge is distinctly higher compared to other hyperparameter regions. Interestingly, this effect is absent in the memoryless learner of Fig.~\ref{fig:NavigationTask} (not shown).  Utilizing such dynamical systems phenomena have the potential to improve the efficiency of hyperparameter search.

\subsection{Renewable resource harvesting}

\begin{figure*}
	\begin{center}
		{\vcenteredhbox{
					\includegraphics[draft=False, width=0.45\linewidth]{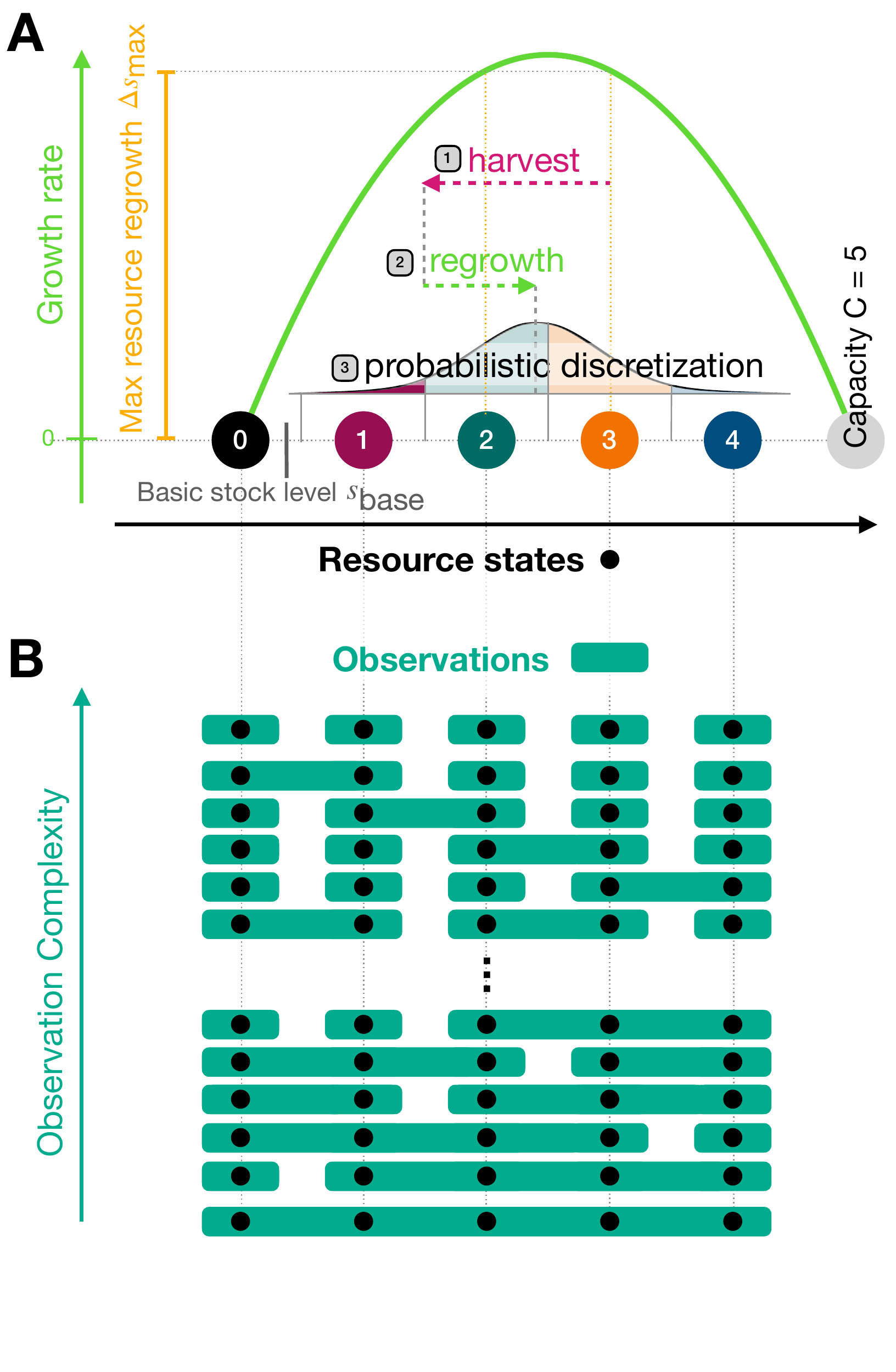}}
			\hspace{0.2cm}
			\vcenteredhbox{
				\includegraphics[draft=False, width=0.5\linewidth]{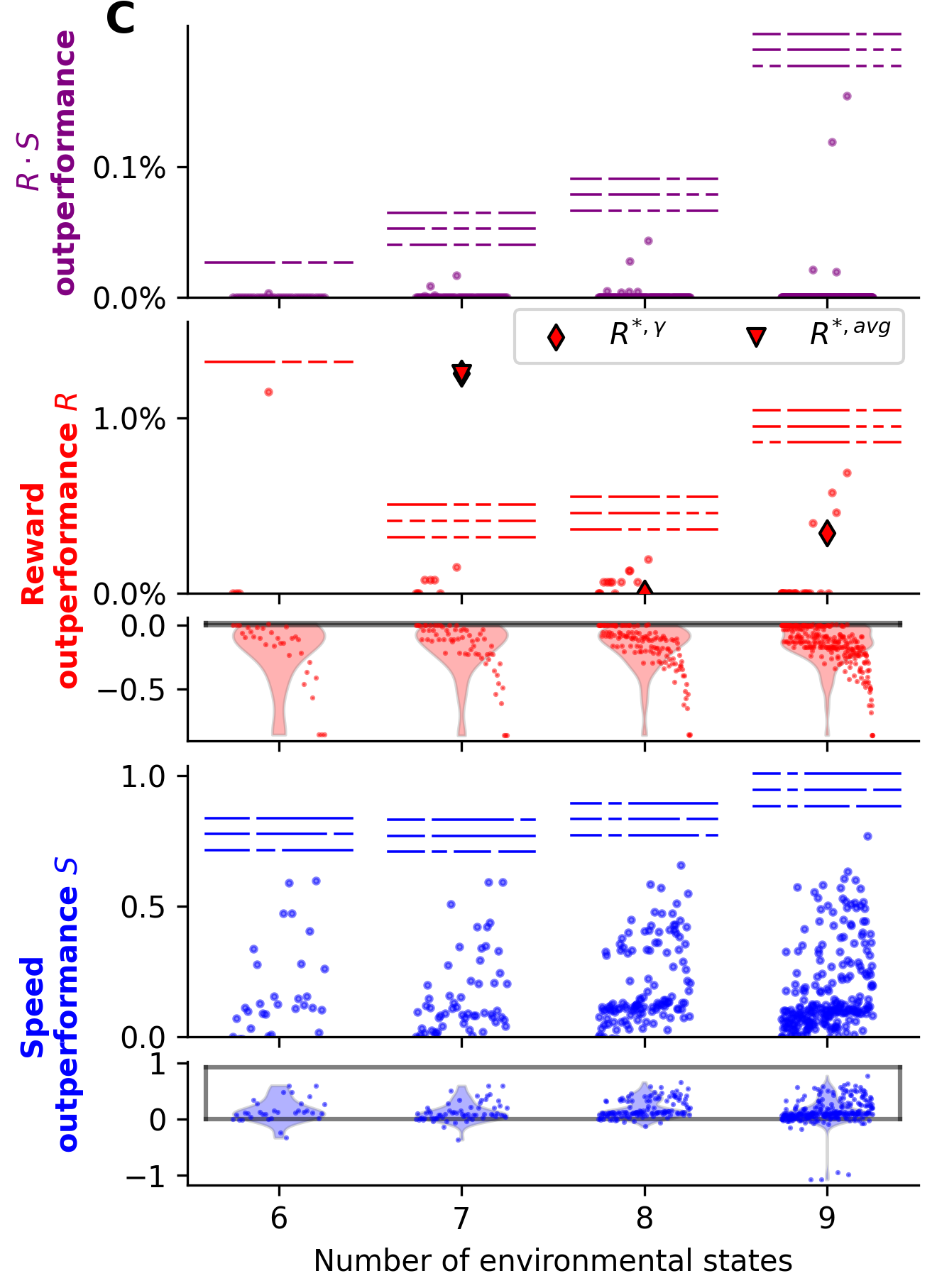}}
		}
		\caption{\textbf{Deterministic learning dynamics in renewable resource environments.} Panel A sketches the functioning of the renewable resource harvesting environment. 1) The agent decides on a harvest, which is subtracted from the current environmental stock. 2) The stock regrows according to a logistic function. 3) The stock is discretized by a normal distribution in order to have the number of states equaling the capacity $C$ of the logistic growth function.
		We set the growth rate $r=0.8$, the effort deviation $\Delta E=0.2$, the stock base level $\tilde s_{base} =0.1$, and the environmental stochasticity $\sigma=0.5$.
		Panel B shows the possible observation spaces -- how the environment is represented by the agent -- ordered by decreasing complexity, for a world in which there are five possible true environmental states. In the most complex (at the top) the agent perceives all real states of the world as distinct; in the least complex (at the bottom), the agent makes the same observation regardless of the true state. We investigate all representations where the agent perceives several adjacent states as a single coherent observation.
		Panels C shows the reward out-performance $R=r/r_\text{ac} - 1$ (red), the speed out-performance $S=1-l/l_{ac}$ (blue), and the combined reward-speed out-performance $R\cdot S (\text{ if } R>0 \land S>0)$ (purple) for all possible representations, for the four renewable resource environments with capacities $C$ and likewise number of states, $6-9$. Out-performance is measured with respect to the agent which used the accurate representation of the environment and obtained a reward $r_\text{ac}$ in $l_{ac}$ time steps.
		For each environment, each dot represents the average of 100 Monte Carlo simulations from random initial policies of a single representation, ordered from the most complex, i.e., the accurate one, on the left to the simplest, i.e., perceiving all states as one, on the right. Violin plots show the distribution of rewards and speed, relative to the agent with the accurate representation.
		The three top performing representations are shown schematically by the dashed lines. Additionally, the average rewards of the optimal discounted policy $R^{*,\gamma}$ and the optimal average-reward policy $R^{*,avg}$ are shown.
		The agent's discount factor $\gamma=0.9$, intensity of choice $\beta'=25$, and learning rate $\alpha=0.02$.
		There exist inaccurate representations (partial observation functions) of the environment that lead to a better learning outcome faster compared to the fully observant agent.}
		\label{fig:ReRe}
	\end{center}
\end{figure*}

\paragraph{Environment description.} Harvesting a renewable resource is a foundational challenge in environmental economics and the Earth and sustainability science \citep{perman2003natural,lindkvist2014modeling,barfuss2017sustainable,geier2019physics}. Here we use a standard logistic growth model, in which the (continuous) resource stock $\tilde s_{t+1} = \tilde s_{t} + r \tilde s_{t} (1- \tilde s_t / C)$ first regrows exponentially with rate $r \in \mathbb R$ until it saturates at capacity $C \in \mathbb N$.
In order to turn the stock-continuous logistic growth  into a state-discrete Markov decision process, we discretize the continuous resource stock into the environmental states $s \in \{0, ..., C-1\}$.
The agent has three possible actions: \textit{harvest nothing}, \textit{harvest a small amount}, or \textit{harvest a large amount}. What is small and large depends on the maximum amount, $\Delta s_\text{max}$, the resource regrows from environmental states $\mathcal S$. The small harvest amounts to $(1-\Delta E) \Delta s_\text{max}$, the large harvest amounts to $(1+\Delta E) \Delta s_\text{max}$, with $\Delta E$ representing the deviation in the agent's harvesting effort.

State transitions work as follows: The harvest amount is subtracted from the current stock state $s_t$. The stock regrows according to the logistic growth equation, yielding a new hypothetical stock $\tilde s_{t+1}$. In order to avoid the complete depletion of the resource, the minimum hypothetical stock yields a value proportional to a base level $\tilde s_{base}$. Since the agent should have an influence on the regrowth of the resource, $\tilde s_{base}$ is multiplied by $(1+\Delta E)$ if the agent chose to \textit{harvest nothing}, by $(1-\Delta E)$ if the agent chose to \textit{harvest a little}, and by $0$ if the agent chose to \textit{harvest a lot}.
The resource stock is then discretized by a normal distribution around $\tilde s_{t+1}$ with variance $\sigma^2$. The probability mass that lies between stock $s_{t+1}-0.5$ and $s_{t+1}+0.5$ gives the probability to transition to the new state $s_{t+1}$. (For $s_{t+1}=0$ the lower bound is $-\infty$, for $s_{t+1}=C$ the upper bound is $+\infty$.) Thus, $\sigma$ represents the level of stochasticity within the environmental dynamics.

The rewards are identical to the harvest amount. Harvesting a lot yields a higher immediate reward than harvesting a little. Except when the resource is degraded, i.e., either the current state $s_t$ or the next state $s_{t+1}$ equals zero, then the rewards are only 10\% of the harvest amount. Thus, the agent has always an immediate incentive to harvest more over a little. The optimal policy depends on the weight the agent puts on future rewards (by its discount factor $\gamma$).

We use this environment to showcase how partial observability can be used to investigate the effect of different (imperfect) representations of the environment. We focus on representations under which the agent perceives several adjacent states as a single coherent observations.
Figures~\ref{fig:ReRe}\,A \& B illustrate the renewable resource harvesting environment and the investigated observation representations for capacity $C=5$.

\paragraph{Results.}
We find that inaccurate (reduced complexity) representations of the environment can lead to a better learning outcome faster, when compared to an agent which perceives the environment accurately (at full complexity) (Fig~\ref{fig:ReRe}).

In the majority of cases an inaccurate representation of the environment leads to a speed out-performance in the order of 10\%, i.e., a smaller number of time steps it takes the learner to converge to a fixed point.
Only four representations of the 9-state environment take distinctly longer to converge. Overall, there is a slight tendency that simpler representations lead to faster convergence. Representations (dots) are ordered from the most complex, i.e., the accurate one, on the left to the simplest, i.e., perceiving all states as one, on the right (per environment).  All top speed representations (dashed bars) cluster the resource stock 0 and 1 together but separate between stock 1 and 2.
In the environments with capacity 8 and 9, a resource stock of 2 is represented completely separate by all top speed representations.

In contrast, the majority of inaccurate representations lead to a worse reward at convergence. Clearly visible by the red dots on the right for each environment, the simpler the representation the worse the performance. Nevertheless, a few representations of intermediate complexity lead to a reward out-performance in the order of 1\%. This is remarkable, since \citet{Blackwell1953}'s theorem showed that a rational decision maker cannot improve by an inaccurate representation. Of course, our result does not contradict Blackwell, since we investigate a learning process.

\begin{table}[b]
	\begin{center}
		\begin{tabularx}{\linewidth}{lXXXX}
		    \hline\hline
			\textbf{Env. states} \ \  & \textbf{6} & \textbf{7} & \textbf{8} & \textbf{9}\\ 
			\hline
			\textbf{Reward $R^{*,avg}$ }  &  0.25 & 0.013 & 0.019 & 0.024 \\ 
			\textbf{Reward $R^{*,\gamma}$}  & -0.015 & 0.013 & 0 & 0.003 \\ 
		    \hline\hline
		\end{tabularx}
		\caption{Average reward of the optimal average-reward policy $R^{*, avg}$ and the optimal policy of the discounted reward setting $R^{*, \gamma}$ for the same four renewable resource environments as in Fig.~\ref{fig:ReRe}. Rewards are also transformed in the same way ($R=r/r_\text{ac} - 1$, with $r_\text{ac}$ being the reward the fully observant agent obtained at convergence).
		\label{tab:Rewards}}
	\end{center}
\end{table}


\begin{figure*}
	\vskip 0.2in
	\begin{center}
		\centerline
		{\vcenteredhbox{
				\includegraphics[draft=False, 	width=0.32\linewidth]{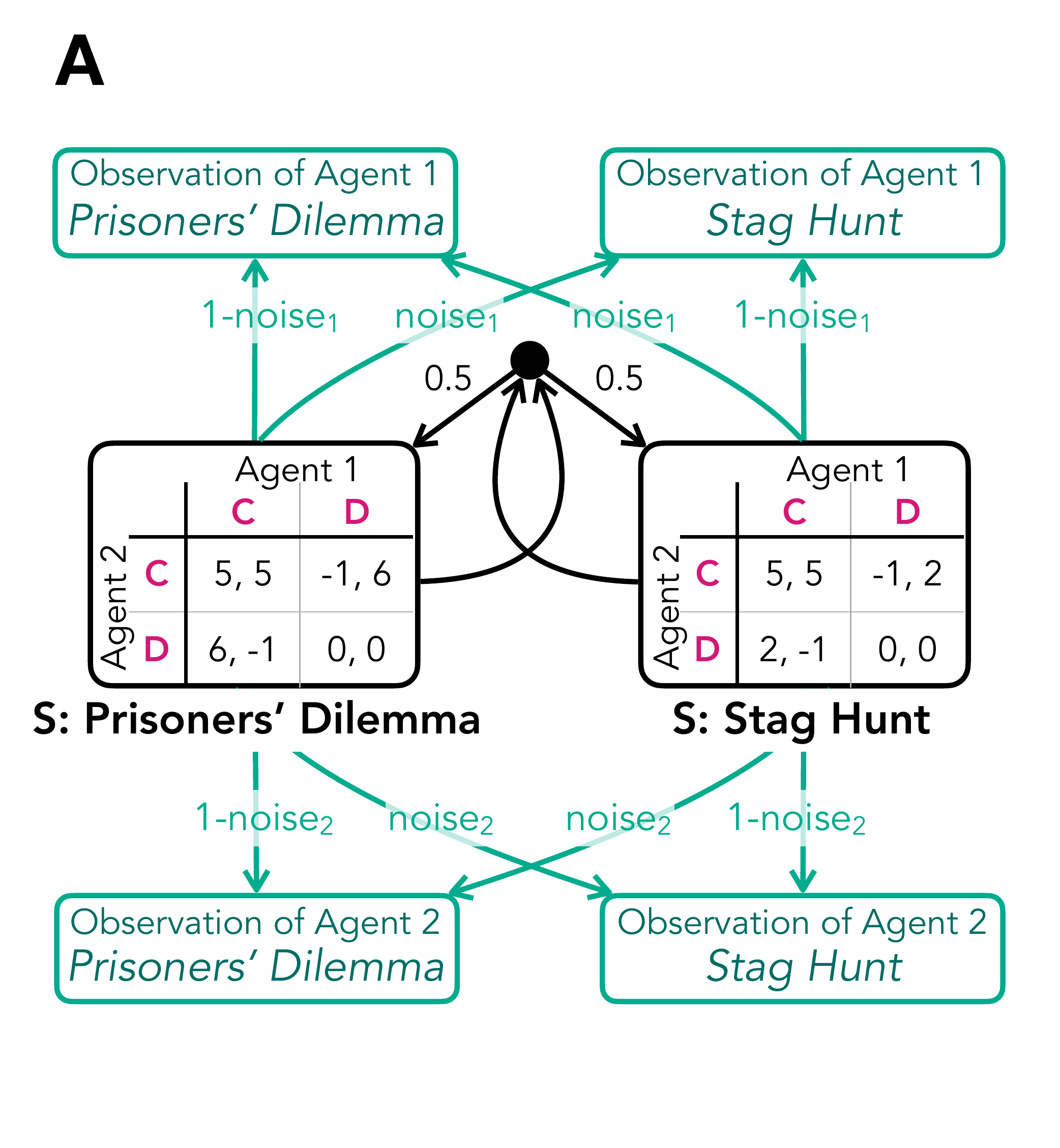}}
			\hspace{0.18cm}
			\vcenteredhbox{
				\includegraphics[draft=False, width=0.65\linewidth]{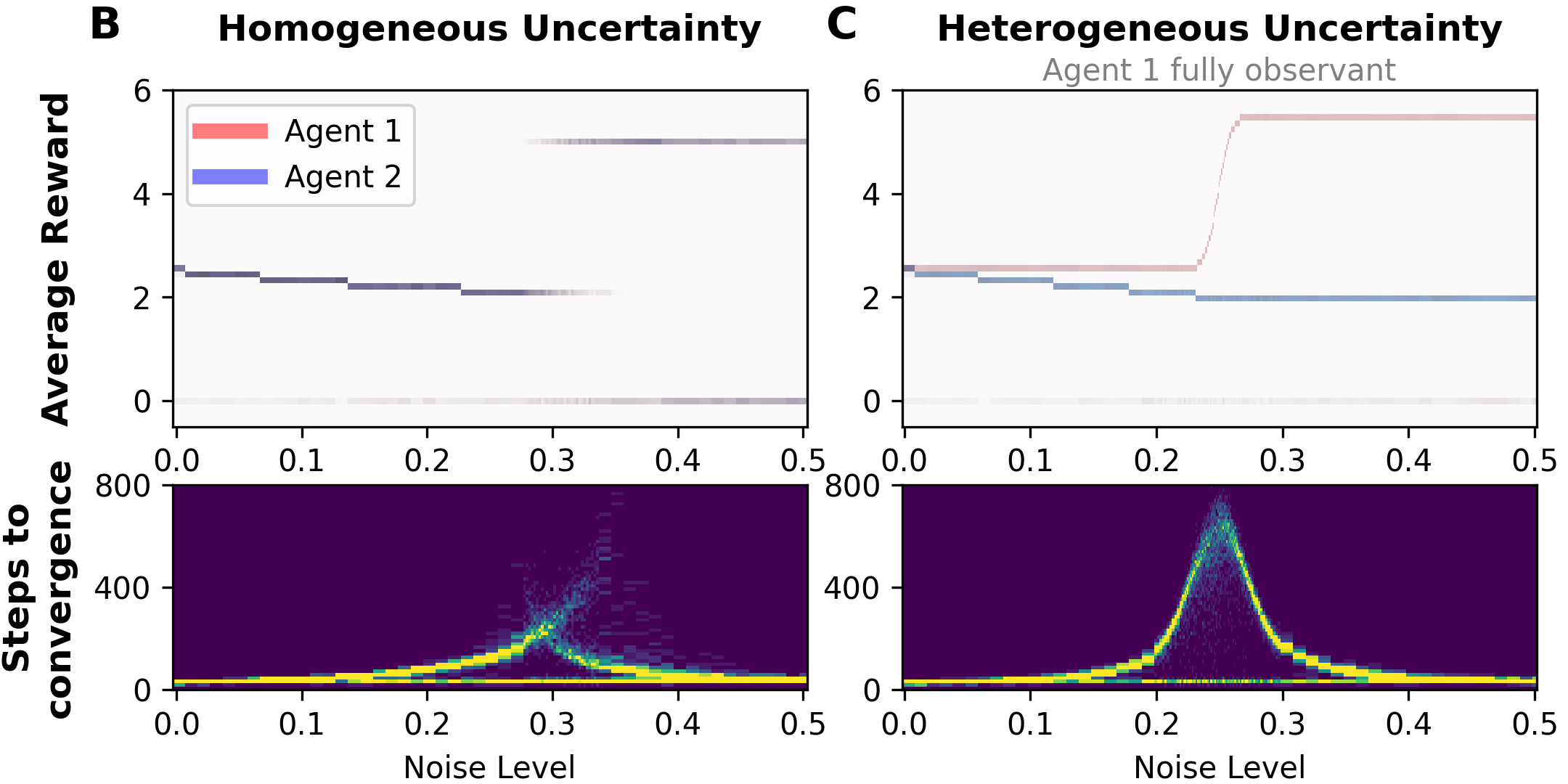}}
		}
		\vspace{-0.25cm}
		\caption{\textbf{Deterministic learning dynamics in an uncertain social dilemma.} Panel A illustrates the environment.
			Panels B and C show the average rewards at convergence for agent 1 in red and agent 2 in blue (top row) and the time steps it takes the learners to convergence (bottom row) for various observational noise levels from 0 to 0.5. For each noise level, the plots show a histogram via the color scale. Each histogram results from a Monte Carlo simulation from 100 random initial policies.
			Panel B shows the case of homogeneous uncertainty where both agents' observations are corrupted equally by noise. In Panel C only agent 2 is increasingly unable to observe the environment correctly (Heterogeneous Uncertainty).
			The discount factor was set to $\gamma = 0.5$ since future states are independent of the agents’ actions, which makes the discount factor irrelevant for the learning in this case. Remaining hyperparameters were set to $\alpha=0.01$ and $\beta' = 50$.
			Homogeneous uncertainty can overcome the social dilemma through the emergence of a stable, mutually high rewarding fixed point above a critical level of observational noise. Heterogeneous uncertainty, however, leads to reward inequality. In both cases, the transition is accompanied by a critical slowing down of the convergence speed.
		}
		\label{fig:UncertainDilemma}
	\end{center}
	\vskip -0.34in
\end{figure*}

To better understand the relationship between the learning process and the rational optimal policies,
Table~\ref{tab:Rewards} shows the average reward of the optimal policy $R^{*,\gamma}$ and the average-reward optimal policy $R^{*,avg}$ relative to the reward obtained by the fully observant agent (shown in Fig~\ref{fig:ReRe} by diamonds and down-triangles). The optimal policy maximizes the state values for each state and depends on the discount factor $\gamma$. The average-reward optimal policy maximizes the average reward. Since in this environment $R^{*,\gamma}$ approaches $R^{*,avg}$ under $\gamma \rightarrow 1$, the rewards between $R^{*,avg}$ and $R^{*,\gamma}|_{\gamma=0.9}$ represent the rewards more patient or future caring agents could obtain.
Thus, the out-performing representations cause the learner to behave as if it were more patient or future-oriented than it actually is (defined by its discount factor $\gamma$). However, it is not obvious to identify regularities across the environments between the top rewarding representations.
Moreover, Table~\ref{tab:Rewards} shows that the learning process under full observability yields decent results. For the environment with 8 states the learner obtains the exact same reward as the optimal policy. In the environment with 6 states the learner obtains an average reward which is even above the one of the optimal policy.

Taken together, a speed out-performance in the order of 10\% multiplied by a reward out-performance in the order of 1\% leads to combined speed-reward out-performance in the order of 0.1\%. Along the four environments investigated, the magnitude in out-performance is increasing with the number of environmental states. Future work is needed to investigate this effect in larger, more complex resource harvesting environments and also how to obtain those representations which lead to better outcomes faster.

Notably, this result resembles the one by \citet{mark2010natural} who show also that simpler views on the world can be of advantage. However, in their model perceiving the truths comes with a cost which is subtracted from the rewards of the environment. If this cost parameter is sufficiently large, then perceiving the truths cannot pay off by design.  We do not model such a cognitive cost of being close to the truths and still find that some inaccurate representations lead to better outcomes faster.

\subsection{Uncertain social dilemma}
\label{sec:Dilemma}

\paragraph{Environment description.}
The emergence of cooperation in social dilemmas is another key research challenge for evolutionary biology and the social and sustainability sciences \citep{nowak2006five,kollock1998social,barfuss2020caring}. We will focus on the situation where two agents can either cooperate (C) or defect (D) and either face a Prisoner's Dilemma or a Stag Hunt game with equal probability \citep[Fig.~\ref{fig:UncertainDilemma}\,A, cf. Refs.][]{LevinePonssard1977,LiCalziMuehlenbernd2019}. In the pure Prisoner's Dilemma defection is the Nash equilibrium, which leads to a suboptimal reward for both agents, also known as the tragedy of the commons \citep{Hardin1968}. In the pure Stag Hunt game, both mutual cooperation and mutual defection are Nash equilibria with the difference that mutual cooperation yields a higher reward than mutual defection for both agents. It is therefore also referred to as a coordination challenge \citep{BarrettDannenberg2012}. Here, we consider the situation when the agents are uncertain about the type of game they are facing at each decision point. Whether we are facing a tragedy or a coordination challenge is relevant for, e.g., the mitigation of human-caused climate change \citep{BarrettDannenberg2017}.
We investigate two scenarios. Under homogenous uncertainty (Fig.~\ref{fig:UncertainDilemma}\,B), both agents' observations are blurred by an increasing level of observational noise. Under heterogeneous uncertainty (Fig.~\ref{fig:UncertainDilemma}\,C), only agent 2's observations become noisier. Since the environment is symmetric under exchanging the roles of the agents, it suffices to explore only one heterogeneous uncertainty scenario.

\begin{figure*}
	\vskip 0.2in
	\begin{center}
		\includegraphics[draft=False, width=0.99\linewidth]{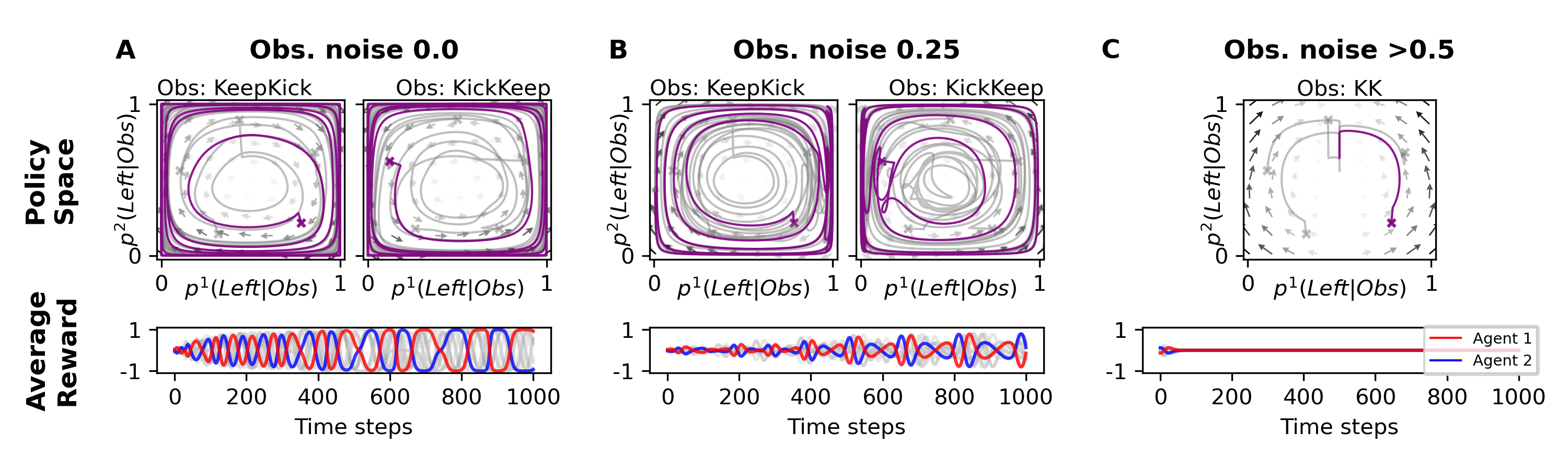}
		\caption{\textbf{Deterministic learning dynamics in an uncertain zero-sum competition.}
			Policy spaces and reward trajectories are shown for three different observational noise levels: (A) $\nu=0.0$, i.e., perfect observation, (B) $\nu=0.25$, and (C) $\nu > 0.5$, i.e., both states are observed inseparably as one. The probability of choosing action \textit{left}, conditioned on the current observation, is plotted on the $x$-axis for agent 1 and on the $y$-axis for agent 2.
			Learning trajectories are shown from five initial policies around the center of the policy spaces. For better visual inspection only one of those trajectories is portrayed in color.
			Arrows in gray indicate the flow of the learning dynamical system.
			Hyperparameter were $\alpha=0.005$, $\beta'=200$, and $\gamma=0.9$.
			Here partial observability is able to stabilize the learning process.
		}
		\label{fig:ZeroSum}
	\end{center}
	\vskip -0.2in
\end{figure*}

\paragraph{Results.}
Homogeneous uncertainty can overcome the social dilemma through the emergence of a stable, mutually high rewarding fixed point above a critical level of observational noise.
Under perfect observation both agents convergence to full defection when observing the Prisoner's Dilemma. When observing the Stag Hunt game it depends on the initial joint policy whether the agents converge to mutual defection or mutual cooperation. Reward values are as such that the defective basin of attraction is comparable small (see the light line at an average reward of $0$ in Fig.~\ref{fig:UncertainDilemma}\,B).
Increasing the observational noise level from zero under homogeneous uncertainty will first decrease the average reward at convergence. The agents still converge to the perfect observation policy which leads them to defect when they observe the Prisoners' Dilemma but the situation is actually the Stag Hunt.
However, increasing observational noise further eventually leads to a bifurcation (Fig.~\ref{fig:UncertainDilemma}\,B). Mutual cooperation under both observations becomes a stable fixed point. As a consequence both agents obtain an average reward of 5 at convergence.
Interestingly, there seems to be a small range of observational noise at which all three rewards $0$, $\sim 2$,  and $5$ are supported by equilibria. For large noise levels only the rewards at 0 and 5 are stable.

Thus, we find that the deterministic learning dynamics under homogeneous partial observability are able to converge to mutually more rewarding policies compared to the perfect observation case. The existence of those equilibria is long known in traditional static game theory \citep{LevinePonssard1977}. Here we show that our derived dynamics are able to serve as a dynamic micro-foundation for those static equilibria. They correspond to fixed points of the derived learning dynamics, and the transitions between equilibria are again accompanied by the dynamical systems phenomenon of a critical slowing down of the convergence speed (Fig.~\ref{fig:UncertainDilemma}\,B, bottom).

However, the mutual benefit of uncertainty vanishes when not all agents' observations are uncertain (Fig.~\ref{fig:UncertainDilemma}\,C). Under slight uncertainty only the reward of the ill-informed agent (Agent 2 in Fig. \ref{fig:UncertainDilemma}) decreases. After the bifurcation point under large uncertainty, the ill-informed agent converges to full cooperation under both observations, whereas the well-informed agent still defects in the Prisoner's Dilemma which earns it an average reward of even more than 5. The knowledgable agent exploits the ill-informed and heterogeneous uncertainty leads to reward-inequality between the agents.

Interestingly, Fig.~\ref{fig:UncertainDilemma} suggests a difference in the type of phase transition between the policy of mediocre reward at low observational noise levels and the policies at high noise levels. The phase transition under homogeneous uncertainty seems to be discontinuous and shifted toward greater noise levels whereas the transition under heterogeneous uncertainty seems to be continuous. Investigating the relationship between the learning dynamics, free energy equivalents \citep{barfuss2021dynamical} and phase transitions is a promising direction of future work.

\subsection{Uncertain zero-sum competition}

\paragraph{Environment description.}
The last environment we use as a test bed is a two-agent, two-state, two-action zero-sum competition,
also known as the two-state matching pennies game \citep{HennesEtAl2010}.
It roughly models the situation of penalty kicks between a kicker and a keeper. Both agents can choose between the \textit{left} and the \textit{right} side of the goal. The keeper agent scores one point if it catches the ball (when both agents have chosen the same action), otherwise the kicker agent receives one point.
The two states of the environment encode which agent is the keeper and which one is the kicker. In state \textit{KeepKick} agent 1 is the keeper and agent 2 is the kicker. In the state \textit{KickKeep} it is the other way around. Agents change roles under state transitions, which depend only on agent 1's actions. When agent 1 selects either \textit{left} as keeper or \textit{right} as kicker both agents will change roles. With symmetrical rewards but asymmetrical state transitions, this two-state zero-sum game presents the challenge of coordinating both agents on playing a mixed strategy with equiprobable actions.
Similarly as in Secs. \ref{sec:Coordination} and \ref{sec:Dilemma}, the agents' observations of the environmental states are obscured by a noise level $\nu$.

\paragraph{Results.}
Figure~\ref{fig:ZeroSum} shows how partial observability can stabilize the learning process. When both agents observe the environment perfectly the learning dynamics are prone to be unstable, either unpredictably chaotic or on periodic orbits and limit cycles \citep[Panel A, ][]{BarfussEtAl2019}.
The rewards of agent 1 and 2 are circulating around zero. Under a medium observational noise level of $\nu=0.25$ the learning dynamics are still unstable. Especially the transient dynamics in the policy space (Panel B, on the right) appear strange. The average reward trajectory looks damped compared to the fully observant agents. Increasing the observational noise further such that the agents perceive the two environmental states (\textit{KeepKick} and \textit{KickKeep}) as a single observation (\textit{KK}), is able to stabilize the learning process. Interestingly, the flow of the learning dynamics is separated into two half circles directed at the upper half of the line at which agent 1 chooses both actions with equal probabilities. As shown by the gray arrows, the circled flow is on a fast timescale compared to the movement downward to the center of the policy space (which is not reached here within 1000 time steps). At this downward movement, both agents play the different roles of kicker and keeper in equal amounts, since only agent 1 is responsible for the state transitions. Any advantage agent 2 gains from deviating from the equiprobable policy as kicker is balanced by the same amount of disadvantage agent 2 looses as keeper. Thus, the rewards for both agents quickly stabilize at zero.

\section{Discussion}
\label{sec:Fin}
In this article we analyzed the efficacy of temporal-difference reinforcement learning  under irreducible environmental uncertainty. To do so, we introduced deterministic multiagent reinforcement learning dynamics, in which the agents
are only partially able to observe the true states of the environment.
These dynamics operate in the theoretical limit of an infinite memory batch, and make implicit inference about the true states via Bayes rule and can be well approximated by finite-size batch learning algorithms. This limit allows us to systematically separate the stochasticity of reinforcement learning, resulting from probabilistic environmental dynamics, observations, and decisions, from the environmental uncertainty that originates in the agents' incomplete awareness of the true state space.

Overall, we have shown how these dynamics can serve as a practical, lightweight, deterministically reproducible and robust tool, to systematically study the combined effects of \textit{strategic uncertainty}, \textit{stochastic uncertainty} and \textit{state uncertainty} in collectives of self-learning agents across a wide range of partially observable environment classes.

We have found  a  variety  of  effects  caused  by  partial  observability, yet general conclusion and recommendations cannot be stated, due to the generality of the partially observable agent-environment setting.
Providing agents with only a partial view of the true state of the world might be expected to always result in poorer learning decision-making outcomes. However, we have demonstrated that irreducible environmental uncertainty can instead lead to better learning outcomes, even in a single-agent environment, stabilize the learning process and overcome social dilemmas in multiagent domains.

Furthermore, our method allows the application of dynamical systems theory  to  partially  observable  multiagent  learning.   We have found  that  partial  observability  can  cause  the  emergence of  catastrophic  limit  cycles, within which the agent obtains the worst possible reward.
We also found instances where partial observability induces phase transitions between low and high rewarding regimes accompanied by a critical slowing down of the learning processes. Further, we saw partial observability induced separations of the learning dynamics into fast and slow eigendirections, as well as multistability of the learning process.

\paragraph{Potential applications.}

These results may be of use in technological applications of multiagent reinforcement learning, with respect to training regimes, hyperparameter tuning, and the development of novel algorithms. For example, if agents are able to detect that they entered a slow eigendirection, then they can safely increase their learning rate for a faster convergence.  Or training regimes and hyperparameter search techniques might be on the lookout for a critical slowing down since this can indicate a phase transition toward high rewarding solutions.
With respect to the hyperparameter values required for a decent performance we found across environments that learning with partial observability demands more exploration and less weight on future rewards, compared to fully observant agents. Moreover, the learning with partial observability might depend crucially on the precise combination of the two parameters, whereas without uncertainty both parameters can be tuned fairly independently.
The fast computation speed and visualization capabilities of the deterministic learning dynamics approach might be particular suited for the challenge to engineer interpretable and safety-critical learning systems.

We have shown that whether partial observability in the classic principle of temporal-difference learning is advantageous depends on the specific nature of the environment and its representation \citep[cf. ecological rationality, ][]{todd2012ecological, hertwig2019taming}.
Given that temporal-difference reinforcement learning is a relatively simple and widely effective algorithm, and one which closely matches known features of neurological learning \mbox{\citep{schultz1997neural,DayanNiv2008}}, this points to a potential evolutionary pressure for agents to develop internal models of the world that do not match the true state space of their environment \mbox{\citep[cf. Refs.][]{mark2010natural,hoffman2015interface}}.
The proposed dynamics are therefore a suitable tool to advance theoretical research in cognitive ecology, which studies how animals acquire, retain, and use information within their ecology, evolution and behavior \citep{shettleworth2009cognition, dukas2009cognitive}. Within this area, research has begun to ask how agents' may evolve nonveridical or incomplete representations of the world \citep{mark2010natural, prakash2020fitness, Mann2021}; the dynamic model presented here offers a tool to study the effect of nonveridical representations in greater depth.

We also showed that partial observability can lead to better collective outcomes in the case of social dilemmas. The question for the preconditions of cooperation and sustainable behavior presents an important area for deeper investigation \citep{barfuss2018optimization,strnad2019deep,bak2021stewardship}. Temporal-difference learning is a widespread principle in neuroscience and psychology \citep{DayanNiv2008} and there is indeed evidence that humans use a payoff-based learning rule in social dilemmas \citep{Burton_Chellew_2015}. The topic of uncertainty is of special relevance in the mitigation of the climate crisis through global cooperation agreements \citep{kolstad2007systematic,milinski2008collective,kolstad2011uncertainty,BarrettDannenberg2012,domingos2020timing}. Our results highlight the potential for a systematic investigation of mechanisms that incorporate useful uncertainty \citep{NaxEtAl2018} for learning and adaptive actors. In our examples, the mutual benefit of uncertainty in the social dilemma vanishes when not all agents are likewise ill-informed causing reward-inequality between the agents. This suggests that partial observability as a mechanism for solving social dilemmas may need to be regulated externally (e.g., by authorities that monitor information flow, or as a feature of the environment) rather than something that is likely to be generated as an evolutionary adaptation amongst individuals in competition with each other.

\paragraph{Future directions.}
A promising directions for future work is the integration of \textit{model uncertainty} through an analytical treatment of noisy dynamics \citep[cf. Ref.][]{galla2009intrinsic}. The stochastic noise models the finiteness of a reasonable learning algorithm compared to the theoretical limit of the infinite memory batch of the present dynamics. The challenge is that this problem is ill defined and many reasonable learning algorithms exist. Furthermore exciting is the embedding of representation and generalization dynamics into the nonlinear dynamics of learning, acting and environment to study the principles of advantageous representations.

\subsection*{Code availability}
Python code to reproduce all results is available at \url{https://github.com/wbarfuss/POLD} and archived at \url{https://doi.org/10.5281/zenodo.6361994}.

\subsection*{Acknowledgements}
This work was supported by UK Research and Innovation Future Leaders Fellowship MR/S032525/1 and the German Federal Ministry of Education and Research (BMBF): Tübingen AI Center, FKZ: 01IS18039A.

\footnotesize
\setlength{\bibsep}{4pt plus 0.5ex}

\end{document}